\documentclass[floatfix,%
 reprint,
 amsmath,amssymb,
 aps,
 prmaterials
]{revtex4-2}

\usepackage{graphicx}
\usepackage{dcolumn}
\usepackage{booktabs}
\usepackage{multirow}
\usepackage{bm}

\newcommand{\boldface}[1]{\boldsymbol{#1}}  

\newcommand{\bfn}{\boldface{n}}

\newcommand{\bfp}{\boldface{p}}
\newcommand{\bfq}{\boldface{q}}

\newcommand{\bfz}{\boldface{z}}
\newcommand{\bfnull}{\boldsymbol{0}}

\newcommand{\bfSigma}{\boldsymbol{\Sigma}}

\newcommand{\calZ}{\mathcal{Z}}

\newcommand{\Rset}{\mathbb{R}}

\newlength{\boxwidth}
\def\dd{\;\!\mathrm{d}}

\def\btheorem{\begin{theorem}}
\def\etheorem{\end{theorem}}
\def\blemma{\begin{lemma}}
\def\elemma{\end{lemma}}
\def\bproposition{\begin{proposition}}
\def\eproposition{\end{proposition}}
\def\bcorollary{\begin{corollary}}
\def\ecorollary{\end{corollary}}
\def\bdefinition{\begin{definition}}
\def\edefinition{\end{definition}}
\def\bexample{\begin{example}}
\def\eexample{\end{example}}
\def\bremark{\begin{remark}}
\def\eremark{\end{remark}}

\newcommand{\be}{\begin{equation*}}
\newcommand{\ee}{\end{equation*}}

\newcommand{\beq}{\begin{eqnarray*}}
\newcommand{\eeq}{\end{eqnarray*}}
\newcommand{\bem}{\begin{multline}}
\newcommand{\eem}{\end{multline}}
\newcommand{\ba}{\begin{align*}}
\newcommand{\ea}{\end{align*}}

\newcommand{\itbf}[1]{\textit{\textbf{#1}}}

\begin{document}

\preprint{APS/123-QED}

\title{Long-term atomistic finite-temperature substitutional diffusion 
}

\author{Shashank Saxena}
\affiliation{Mechanics \& Materials Lab, Department of Mechanical and Process Engineering, ETH Z\"urich, 8092 Z\"urich, Switzerland}

\author{Prateek Gupta}
\affiliation{Department of Applied Mechanics,  Indian Institute of Technology Delhi, 110016, New Delhi, India}

\author{Dennis M. Kochmann}
\email{dmk@ethz.ch}
\affiliation{Mechanics \& Materials Lab, Department of Mechanical and Process Engineering, ETH Z\"urich, 8092 Z\"urich, Switzerland}

\date{\today}

\begin{abstract}
Simulating long-term mass diffusion kinetics with atomic precision is important to predict chemical and mechanical properties of alloys over time scales of engineering interest in applications, including (but not limited to) alloy heat treatment, corrosion resistance, and hydrogen embrittlement. We present a new strategy to bridge from the time scale of atomic vibrations to that of vacancy-mediated atomic hops by a combination of
statistical mechanics-based Gaussian phase packets (GPP) relaxation and a nudged elastic band (NEB)-facilitated harmonic transition state theory (H-TST) time update. We validate the approach by simulating bulk self-diffusion in copper and the segregation of vacancies and magnesium to a stacking fault and a symmetric tilt grain boundary in aluminum, modeled with an embedded atom method (EAM) potential. The method correctly predicts the kinetics in bulk copper and equilibrium impurity concentrations in aluminum, in agreement with the Langmuir-Mclean solution in the dilute limit. Notably, this technique can reach realistic diffusion time scales of days, weeks, and even years in a computational time of hours, demonstrating its capability to study the long-term chemo-thermo-mechanically coupled behavior of atomic ensembles.

\end{abstract}

\maketitle

\section{Introduction} \label{intro}

Accurate knowledge of the long-term diffusion behavior in alloys is essential for the rational design of both alloy compositions and microstructures. However, the experimental determination of diffusion coefficients is usually possible only at elevated temperatures, because diffusion rates are prohibitively low at low temperatures, making them difficult to measure within practical time frames and often falling below detection thresholds \citep{sauer1962diffusion, paul2014thermodynamics}. As a result, diffusion data are typically obtained from high-temperature experiments and then extrapolated to lower temperatures. This approach, however, introduces significant uncertainties due to changes in diffusion mechanisms or defect structures with temperature \citep{kaur1995grain}. Furthermore, traditional tracer diffusion and interdiffusion experiments provide bulk-averaged values, which may not reflect the local diffusion behavior near microstructural features such as grain boundaries or dislocations, regions that are often critical to material performance \citep{DIVINSKI20073337}. These limitations underscore the inadequacy of relying solely on experimental data, e.g., for designing high-performance alloys. To bridge this gap, there is a need for modeling techniques that can predict diffusion behavior at microstructural length scales using the interatomic forces, while enabling the study of concentration evolution over technologically relevant time scales.

Macroscopic diffusion is usually modeled by Fick's law of diffusion in solids and fluids. For anisotropic crystals, Onsager showed that, using microscopic reversibility of atomic collisions, symmetry relations between the so-called Onsager coefficients coupling multiple transport processes can be obtained~\citep{PhysRev.37.405}. Using random walk models, the diffusion coefficient in Fick's law can be related to the mean-squared displacement of a diffusing particle in free space, using the Einstein-Smoluchowski relation \citep{ESR}. If the particle is diffusing in a crystal lattice, then the mean-squared displacement also involves correlations between different hops~\citep{mehrer2007diffusion}. To obtain the hopping rate of individual atoms, Vineyard~\cite{vineyard1957frequency} expressed it in terms of the total number of configurations close to equilibrium states in the $N$-dimensional potential landscape and the number of configurations close to the $(N-1)$-dimensional manifold passing through the saddle point between the equilibrium states. These hopping rates can be used in a master equation, which governs the evolution of concentrations at atomic sites. \citet{Nastar01012000} used a parameterized configurational Hamiltonian description to evaluate approximate kinetic rates of configurational changes of a crystal. However, all of the above works consider diffusion in bulk crystals, focusing on \textit{bulk} diffusion coefficients in alloys and are not applicable to local diffusion in the vicinity of defects.


Numerical techniques to simulate mass diffusion in atomic ensembles can be classified into two major categories: (i) trajectory-following algorithms and (ii) statistical ensemble averaging algorithms. Conventional molecular dynamics (MD) belongs to the former category and has been used with a mean square displacement (MSD) analysis to compute diffusion constants in alloys such as for InGaN \citep{ZHOU2017331}, CuNi \citep{KHODAKARAMI2020107712}, and ZrNi \citep{AIHARA1996201} as well as in pure metals \citep{ROSSI199127}. Green-Kubo velocity autocorrelation function (VACF) analysis \citep{PhysRevA.6.1214} has also been used to study the diffusion process. Unlike MSD, VACF captures time-dependent correlation effects and can resolve subtle collective motion mechanisms like vortex formation \citep{PhysRevB.111.L081104} and cage dynamics \citep{BADIA20171}. Traditional MD is limited by the need to resolve fast thermal vibrations, which limits the time scales accessible by MD studies. Numerous accelerating approaches have been proposed to advance the MD trajectories to longer time scales. These include, but are not limited to, hyperdynamics \citep{10.1063/1.473503, voter1997hyperdynamics}, parallel replica dynamics \citep{PEREZ201590}, and collective variable-driven hyperdynamics (CVHD) \citep{CVHD, FUKUHARA2020109581, EBINA2021110577}. Kinetic Monte Carlo (KMC) \citep{voter2007introduction} is another paradigm of trajectory-following simulations, which focuses on the configurational dynamics instead of the vibrational dynamics. These simulations can achieve time scales much longer than MD and are often combined with ab-initio-computed transition rates~\citep{PhysRevMaterials.4.103601, PhysRevMaterials.2.123403, PhysRevMaterials.5.013803, tafen2015first}. Traditional KMC approaches require a pre-identification of the possible transition pathways \citep{PhysRevB.34.6819, voter1988simulation}, which is often referred to as the \textit{rate catalog}. Unfortunately, this is practical and accurate only for a bulk lattice. Near defects, the vibrational spectrum of atoms affects the transition probabilities \citep{TINGAUD2010727}, and different configurations have different vibrational atomic spectra \citep{herrmann2007finding}. Hence, KMC has often been combined with cluster MD \citep{vzenivsek2008combined, TAVENNER2023111929} to replicate such effects, which makes the entire approach computationally cumbersome and expensive. Machine learning of transition probabilities has also been applied to speed up KMC simulations \citep{atoms9010002}. Traditional lattice-based KMC techniques rely on a pre-defined lattice, hence eliminating the need for spatial discretization. Therefore, their application is limited to systems in which the local microstructure remains fixed despite the presence of local stress inhomogeneities. To bypass this limitation, off-lattice KMC techniques have been developed, in which the physical space is discretized according to the event catalogue~\citep{ruzayqat2018rejection}. However due to full spatial discretization, such techniques are limited to a small size systems. Consequently, there is a need for a modeling technique that can simulate diffusion at atomistic scales for systems sufficiently large to accommodate crystal defects, while at the same time accounting for local microstructural changes due to mass diffusion.

Numerical techniques based on statistical ensemble averaging postulate a form of the probability density function (PDF), usually of Gaussian type, for the atomic ensemble and track the evolution of the parameters of this PDF over time \citep{ma1993approximate, gupta2021nonequilibrium,romero2023extended}. This is accomplished by recourse to Liouville's equation \citep{zubarev1973nonequilibrium} and is equivalent to the maximum-entropy (\textit{max-ent}) formalism \citep{kulkarni2008variational}  in the quasistatic limit. Extending this technique to multiple chemical species is achieved through the concept of atom-wise chemical concentrations \citep{SANCHEZ1984334}. The evolution of chemical concentrations by using Onsager's linear kinetics has been termed \textit{diffusive molecular dynamics} (DMD) and applied to many applications of mass transport \citep{li2011diffusive, ponga2018unified, sun2017acceleration, sun2019atomistic, sun2024exploring, dontsova2014solute, dontsova2015solute, venturini2014atomistic}. However, these techniques require a continuum-level diffusivity constant fitted to experimental values of bulk diffusion. Hence, their ability to accurately capture mass diffusion kinetics at the atomic scale near defects is limited. Consequently, there exists a need to develop computational techniques that can: (i)~predict the dynamic concentration evolution of solutes in an alloy in the presence of lattice defects, (ii)~operate solely based on a given interatomic potential without the need for phenomenological Onsager coefficients, and (iii) reach time scales of engineering interest. This work aims to address that need.

We combine harmonic transition state theory (HTST) \citep{vineyard1957frequency} with local-environment nudged elastic band (NEB) \citep{henkelman2000improved} relaxations to obtain accurate transition rates for an atomic ensemble at finite temperature, with thermo-mechanical relaxation governed by a Gaussian phase packets (GPP) framework. This ensures a better estimation of the environment and temperature-dependent energy barriers \citep{TDEB} than those from atomic positions from ensembles relaxed at $0~$K using molecular statics, and a computationally efficient estimation of the jump frequency by using the GPP-relaxed position variances. Moreover, we also present a procedure to obtain vacancy binding energies required for intermetallic substitutional diffusion \citep{PhysRev.56.814} on-the-fly. To make the approach computationally efficient, we choose only unique atomic environments for computing the transition rates and vacancy binding energies by using basic atomic environment descriptors.

The remainder of this contribution is structured as follows. Sect.~\ref{sec: theory} summarizes the basic theory of statistical mechanics relevant to the GPP framework to obtain thermally relaxed atomic ensembles at finite temperature, the basic concepts of a master equation to model Markovian processes, and the role of vacancy binding energies and \textit{dilute-vacancy approximation} in intermetallic diffusion to reach long time scales. Sect.~\ref{sec: methods} describes the numerical methods used to estimate transition energy barriers and vacancy binding energies for unique atomic environments, along with a crucial \textit{dilute-vacancy} approximation to reach long time scales for intermetallic diffusion. Next, Sect.~\ref{sec: examples} presents the numerical results in two parts. First, we demonstrate the validity of the method to capture self-diffusion in copper and aluminum by comparing to experimental data and the Langmuir-McLean isotherm \citep{mclean1957grain}, respectively. Second, we showcase the ability of the method to follow the long-term kinetics of vacancy-mediated magnesium segregation to a stacking fault and a symmetric tilt grain boundary in aluminum. Finally, Sect.~\ref{sec: conclusion} 
concludes this study and discusses future directions.

\section{Theoretical Background} \label{sec: theory}

\subsection{Statistical atomic representation}
We consider an ensemble of $N$ interacting atoms (labeled by subscripts $i=1,\ldots, N$) of $M$ distinct chemical species (indicated by superscripts $\alpha=A,B\dots$) with vacancies ($v$) that facilitate the substitutional movement of atoms. The distribution of species within the ensemble at time $t$ is described by the set $\bm n(t) = \{\bm n_i(t) : i=1,\ldots,N  \}$ of all site vectors $\bm n_i$, whose element $n_i^k$ is equal to 1 if site $i$ is occupied by species $k$ and equal to 0 otherwise. Superscript $k$ ($\in \{v,A,B\dots\}$) represents any of the $M$ chemical species or a vacancy. Moreover, every site must be occupied by an atom or be vacant. Therefore,
\begin{align}
    \sum_k n_i^k = n_i^v + \sum_\alpha n_i^\alpha = 1  \qquad \forall \ i=1,\ldots,N.
\end{align}

An instantaneous position of the ensemble in phase space is described by the set of all atomic positions $\bfq (t)=\{\bfq_i(t) : i=1,\ldots,N\}$ and momenta $\bfp (t)=\{\bfp_i(t) : i=1,\ldots,N\}$, which can be written as a set of condensed phase-space coordinates: $\bfz=\left(\bfp(t),\bfq(t)\right)  \in \mathbb{R}^{6N}$. Hence, the configuration vector $\bm n(t)$ and the phase space coordinate $z(t)$ completely determine an instantaneous state of the system. Conventional MD simulations are suitable for studying the evolution of the instantaneous state of the system. However, they are limited to short time scales, making the long-term non-equilibrium evolution of the system at the time scales of typical engineering interest inaccessible by MD. As a remedy, we resort to statistical mechanics and assume an interparticle-independent probability density function of the ensemble,
\be
  f(\bm n, \bfz ,t) = \prod_{i=1}^N f_i(\bm n_i,\bfz _i,t)
\ee
with
\be
\begin{split}
& f_i(\bm n_i,\bfz _i,t) = \\ 
& \quad \frac{ \bm c_i \cdot \bm n_i } {\calZ _i(t)} \exp\left[ -\frac{1}{2}(\bfz _i - \bar{\bfz _i}(t))^T \boldsymbol{\Sigma}_i^{-1}(t) (\bfz _i - \bar{\bfz _i}(t))\right] ,
\end{split}
\ee
where $\bar{\bfz }_i(t)= \langle\bfz_i\rangle= \sum_{\bm n} \int f_i(\bm n,\bfz,t)\bfz\dd\bfz$ denotes the mean phase space coordinate, $\bfSigma_i\in \mathbb{R}^{6\times 6}$ is the covariance matrix of the positions and momenta of atom~$i$, and $c_i^k = \langle n_i^k \rangle$ is the phase-space-averaged concentration of species $k$ at site~$i$. Note that $ \sum_k c_i^k = c_i^v + \sum_\alpha c_i^\alpha =1$ for all atomic sites. The partition function $\calZ_i(t)$ is defined such that $ \sum_k \int_{\bfz} f_i(\bm n_i,\bfz_i,t) \text{d}\bfz_i = 1$ where $k-$th component of $\bm n_i$ is 1 and others are 0. The long-term evolution of substitutional mass diffusion in the atomic ensemble is achieved by tracking the evolution of the parameters of the probability density function, $( \{\bar{\bfz}_i \}, \{\boldsymbol{\Sigma}_i \}, \{\bm c_i \} )$. To this end, we assume a separation of time scales between the statistical relaxation of atomic hops and those of atomic vibrations. In other words, we assume that the convergence of configurational rearrangements (described by the set of concentrations $\bm c$) happens on a time scale much longer than the thermo-mechanical relaxation (described by the set of mean positions and variances). Consequently, at any given time of the non-equilibrium concentrations $\{\bm c\}$, we assume that the mean positions and position variances $\{ \bar{\bm z}_i, \Sigma_i \}$ equilibrate instantaneously.

\subsection{Relaxed atomic structures}
Based on the aforementioned separation of time scales, we need to compute relaxed atomic structures and position variances for a given fixed concentration field $\{\bm c_i\}$ of the ensemble. To this end, we utilize the independent Gaussian Phase Packets (GPP) formalism of \citet{gupta2021nonequilibrium}, which is reviewed here briefly to the extent necessary for this work. The GPP framework assumes a hyperspherical atomic distribution function $f_i(\bm n_i,\bm z_i,t)$ in the six dimensions spanned by positions and momenta of atom~$i$. This simplifies the covariance matrix, which may be expressed through block matrices as
\begin{align}
     \bfSigma_{i} = \left(\begin{matrix}
                    \bfSigma^{(\bfp,\bfp)}_{i} &  \bfSigma^{(\bfp,\bfq)}_{i} \\
                     \bfSigma^{(\bfq,\bfp)}_{i} &  \bfSigma^{(\bfq,\bfq)}_{i}
                       \end{matrix} \right) = 
                   \left(\begin{matrix}
                    \Omega\, \bf I &  \beta \, \bf I \\
                     \beta \, \bf I &  \Sigma \, \bf I
                       \end{matrix}
\right),
\label{gpp matrix details}
\end{align}
where $\bf I$ is the identity matrix in three dimensions and $\Omega,\beta,\Sigma\in\Rset$ characterize the (isotropic) momentum-momentum, momentum-position, and position-position covariances of atom~$i$, respectively. Hence, the set of parameters to solve for at every atomic site $i$ becomes $(\bar{\bm q}_i, \bar{\bm p}_i, \Sigma_i, \Omega_i, \beta_i  )$. At thermodynamic equilibrium, the mean momenta $\bar{\bm p_i}$ and thermal momenta $\beta_i$ vanish for every atom, and momentum variances $\Omega_i$ are defined by the process which brings the system to equilibrium \citep{gupta2021nonequilibrium}. In this work, we focus on isothermal conditions and a uniform temperature, for which $\Omega_i = m_i k_B T$ for every atomic site in the ensemble, with $m_i$ being the mass of atom $i$, $k_B$ Boltzmann's constant, and $T$ the absolute temperature. The quasistatic limit of Liouville's equation applied to the probability distribution function further yields the nonlinear equations to be solved for the set of average positions $\bar\bfq=\{ \bar\bfq _i: i=1,\ldots,N \}$ and position variances $\Sigma=\{\Sigma_i : i=1,\ldots, N\}$ for all atoms in equilibrium:
\begin{align}
&\langle \itbf{F}_i \rangle  = \bfnull\qquad \mathrm{and} \nonumber \\
&\frac{\Omega_i}{m_i} + \frac{ \langle \itbf{F}_i(\bm q ) \cdot (\bm q  - \bar{\bm q })  \rangle }{3}  = 0,\label{eq: EOM QS}
\end{align}
The average force on an atom $i$ can be written as $\langle \itbf{F}_i \rangle = -\partial \langle V(\bm q, \bm n) \rangle/ \partial \bar{\bm q}_i$, where $V(\bm q, \bm n)$ is the total potential energy of the system as a function of the atomic positions and chemical species. This study uses interatomic potentials to define the total potential energy. Solving Eq.~\eqref{eq: EOM QS} requires the evaluation of phase averages of $V( \bm q, \bm n )$ and its derivatives. The phase average of the potential energy can be written as
\begin{align}
    \left\langle V (\bm q, \bm n)  \right\rangle  &= \int_{\Gamma} \sum_{\bm n}   V(\bm q, \bm n)  \prod^{n_N}_{j=1} \left(\bm c_j\cdot \bm n_j\right) \text{exp}\left[ -\frac{ \vert \bm q_{j} - \bar{\bm q}_{j} \vert^2}{2 \Sigma_{j} }  \right] \text{d}\bm q_{j} \nonumber \\
    & = \int_{\Gamma} \hat{V}(\bm q, \bm c) \prod^{n_N}_{j=1} \text{exp}\left[ -\frac{ \vert \bm q_{j} - \bar{\bm q}_{j} \vert^2}{2 \Sigma_{j} }  \right] \text{d}\bm q_{j},
\label{phase avg V}
\end{align}
where ${n_N}$ is the number of atoms in the atomic cluster formed by atom $i$ and its interacting neighbors. The approximation assumes the existence of an effective potential $\hat{V}(\bm q, \bm c) = 
\sum_{\bm n}  V(\bm q, \bm n)\prod^{n_N}_{j=1} \left(\bm c_j\cdot \bm n_j\right)$. Finding the exact effective potential is straightforward for a pair potential, while approximate effective potentials exist \citep{PhysRevB.40.10322} for many-body potentials (such as potentials of the embedded atom method (EAM)). In this work, we use the random alloy model proposed by \citet{PhysRevB.93.104201} for EAM potentials.

\subsection{Concentration evolution}
To model the configurational kinetics, we use the master equation for a Markovian process \citep{crispin2009stochastic}, which describes the rate of change of the probability of finding the ensemble in a particular configuration as
\begin{align}
    \frac{\dd P(\bm n,t) }{\dd t} = \sum_{\bm n'} \left[  P(\bm n',t) W(\bm n' \rightarrow \bm n) - P(\bm n,t) W(\bm n \rightarrow \bm n')  \right].
    \label{eq: original master eq}
\end{align}
Here, $P( \bm n,t ) = \int_\Gamma f( \bm n, \bm z, t ) \dd \bm z$ is the marginal probability of finding the system in a configurational state $\bm n$ at time $t$, and $W(\bm n' \rightarrow \bm n)$ is the transition rate of the system from state $\bm n'$ to $\bm n$. Assuming that a change of state is possible only by the swap of an atom with a nearest-neighbor vacancy leads to
\begin{align}
    \frac{\dd c_i^\alpha }{\dd t} = \sum_{j \in \text{NN}^i} \left \langle n_i^v n_j^\alpha \gamma_{j \rightarrow i}^{\alpha} - n_j^v n_i^\alpha \gamma_{i \rightarrow j}^{\alpha}    \right \rangle,
    \label{eq: master eq 1}
\end{align}
where NN$^i$ denotes the nearest-neighbor shell around site $i$ and $\gamma_{i \rightarrow j}^{\alpha}$ is the transition rate of the system for an atom of chemical species $\alpha$ to move from site $i$ to a vacant site~$j$, keeping the configuration of all other sites unchanged. The reader is referred to Appendix \ref{appendix: ME simpl.} for the derivation of Eq.~\eqref{eq: master eq 1} from Eq.~\eqref{eq: original master eq}. It is important to note that, for a given system configuration $\bm n$, the transition rate $\gamma_{i \rightarrow j}^{\alpha}$ depends on the occupancy vectors of all sites other than those of $i$ and $j$. Moreover, the coupled ordinary differential equations in~\eqref{eq: master eq 1} are for all species~$\alpha$ except the vacancies, which are complemented by the constraint $c_i^v + \sum_\alpha c^\alpha_i = 1$ at each atomic site. It is known from the hole theory of diffusion~\citep{PhysRev.56.814} that the presence of an atom of a solute species~$\alpha$ causes the neighboring vacancy concentration in the solvent to change at a time scale significantly faster than that of concentration changes of the solute atom or the surrounding atoms. This is due to different binding energies of a solute-vacancy pair and a solvent-vacancy pair. Consequently, Eq.~\eqref{eq: master eq 1} can be further simplified to
\begin{align}
    \frac{\dd c_i^\alpha }{\dd t} 
    &= \sum_{j \in \text{NN}^i} \left[ \langle n_i^v n_j^\alpha  \rangle \left \langle \gamma_{j \rightarrow i}^{\alpha} \right \rangle - \left \langle n_j^v n_i^\alpha \right \rangle \left \langle  \gamma_{i \rightarrow j}^{\alpha}    \right \rangle\right] \nonumber \\
   &= \sum_{j \in \text{NN}^i} \left [ \kappa^\alpha_{ji} c_i^v c_j^\alpha \langle \gamma_{j \rightarrow i}^{\alpha} \rangle 
    -  \kappa^\alpha_{ij} c_j^v c_i^\alpha \langle\gamma_{i \rightarrow j}^{\alpha}\rangle    \right ],
    \label{eq: approx master eq we use 2}
\end{align}
where $\kappa^\alpha_{ji} = \exp{\left(-F_{\text{int}_{ji}}^{\alpha v} /k_BT\right)}$ is the \textit{vacancy enrichment factor} at site $i$ as a function of the interaction free energy when the solute $\alpha$ is at site $j$ and the vacancy is at site $i$, and analogously for $\kappa^\alpha_{ij}$. In the special case of only vacancies and a single species, there is no free energy difference between two configurations having different atoms near a vacancy, since all atoms are of the same species. Hence, in this case $F_{\text{int}_{ji}}^{\alpha v}=0$ for all nearest neighbor pairs $i$ and $j$. Therefore, Eq.~\eqref{eq: approx master eq we use 2} further simplifies to
\begin{align}
    \frac{\dd c_i^\alpha }{\dd t} = \sum_{j \in \text{NN}^i} \left [ c_i^v c_j^\alpha \langle \gamma_{j \rightarrow i}^{\alpha} \rangle - c_j^v c_i^\alpha \langle\gamma_{i \rightarrow j}^{\alpha}\rangle    \right ].
    \label{eq: master_eq_simplified}
\end{align}

\subsection{Diffusion in intermetallics in the dilute vacancy limit} 
\label{Diffusion in Intermetallics}

To simulate the concentration evolution in alloys (or intermetallics), we must integrate the set of coupled nonlinear ordinary differential equations in Eq.~\eqref{eq: approx master eq we use 2} for all species $\alpha$ except the vacancies for all atomic sites. The time scale of concentration changes governed by Eq.~\eqref{eq: approx master eq we use 2} corresponds to the time scale at which solute and/or solvent atoms replace vacancies. This is the time scale of vacancy diffusion. However, in the dilute vacancy limit ($c_i^v \ll 1$ for all sites $i$), the time scale of solute diffusion is significantly slower than the vacancy diffusion time scale. To access such slow time scales, the time steps in numerical integration of Eq.~\eqref{eq: approx master eq we use 2} must be significantly larger than the vacancy diffusion time scales, which can result in $c_i^v$ becoming negative in numerical simulations. Therefore, we assume that the vacancy concentration relaxes to the movement of solute and/or solvent species almost instantaneously, resulting in the constraint,
\begin{equation}
    c^v_i = c^v_0 \exp\left[-F_{\text{seg}_i}^v/k_BT\right] = c^v_0 \lambda^v_i,
    \label{eq: fast_vacancy_constraint}
\end{equation}
where $F_{\text{seg}_i}^v$ is the segregation free energy of a vacancy at site $i$, $\lambda^v_i$ is the corresponding \textit{vacancy segregation factor}, and $c_0^v$ is the bulk vacancy concentration. Substituting Eq.~\eqref{eq: fast_vacancy_constraint} in Eq.~\eqref{eq: approx master eq we use 2}, we obtain
\begin{align}
    \frac{\dd c_i^\alpha }{\dd t} = c_0^v \sum_{j \in \text{NN}^i} \left [ \kappa^\alpha_{ji} \lambda_i^v c_j^\alpha \langle \gamma_{j \rightarrow i}^{\alpha} \rangle 
    -  \kappa^\alpha_{ij} \lambda_j^v c_i^\alpha \langle\gamma_{i \rightarrow j}^{\alpha}\rangle    \right ].
    \label{eq: approx master eq we use alloys}
\end{align}
Eq.~\eqref{eq: approx master eq we use alloys} allows us to simulate long-term solute diffusion in the dilute vacancy limit, since the fast vacancy diffusion is assumed instantaneous, thereby relaxing the stiffness of the coupled ordinary differential equations in~\eqref{eq: approx master eq we use 2}. Specifically, we numerically integrate Eq.~\eqref{eq: approx master eq we use alloys} for all solute chemical species and find the solvent chemical species using the approximate constraint $\sum_\alpha c^\alpha_i\approx 1$.

Time integration of Eq.~\eqref{eq: approx master eq we use alloys} requires calculation of local environment-dependent vacancy enrichment factors, vacancy segregation factors, and mean transition rates between two atomic sites, which are discussed in detail in the next section.

\section{Methodology}   \label{sec: methods}
At each time step, we use staggered solution scheme, alternating between quasistatic relaxation (to obtain the relaxed mean atomic positions and position variances $\{ \bar{q}_i,\Sigma_i \}$ as a function of the current chemical concentrations $\{\bm c_i\}$ at all sites by solving Eqs.~\eqref{eq: EOM QS}) and an update of the concentrations $\{\bm c\}$ (by solving Eq.~\eqref{eq: master_eq_simplified}).

Solving Eqs.~\eqref{eq: EOM QS} for every atomic site involves the evaluation of a high-dimensional integral over the phase space of positions of all atoms, as shown in Eq.~\eqref{phase avg V}. This is usually achieved by Gaussian quadrature rules \citep{doi:10.1137/1015023}, pre-integrated values from lookup tables \citep{li2011diffusive,dontsova2014solute,dontsova2015solute}, or graph neural networks (GNNs) trained on Monte Carlo data \citep{SAXENA2023104681}. In this work, we use pre-integrated lookup tables for EAM potentials, as shown by \citet{li2011diffusive}.

To update the concentrations of every atomic site and chemical species through Eq.~\eqref{eq: master_eq_simplified}, we must model the average forward and backward transition rates ($\langle \gamma_{i \rightarrow j}^{\alpha} \rangle$ and $\langle \gamma_{j \rightarrow i}^{\alpha} \rangle$, respectively) for every pair of nearest neighbors ($i$ and $j$) and every chemical species in the ensemble. This is performed by using H-TST \citep{vineyard1957frequency} informed by local environment-dependent energy barriers computed using the NEB \citep{henkelman2000improved} technique, as described in the following.

\subsection{ Finding energy barriers } \label{subsec: NEB impl.}
The transition rate for a hop of atomic species $\alpha$ from site $i$ to $j$ depends on the atomic species present at the neighboring sites of both $i$ and $j$. Using the rate expression from H-TST, the average transition rate $\langle \gamma_{j \rightarrow i}^{\alpha} \rangle$ can be written as
\begin{align}
    \langle \gamma_{i \rightarrow j}^{\alpha} \rangle &= \nonumber \\ 
    \sum_{\bm n} &  \left(  \prod_{ \substack{p=1 \\ p\neq i,j}  }^N \left( \bm c_p \cdot \bm n_p \right) \right)
\nu(\bm n, \{ \bar{\bm q} \}) \exp{\left[-E_\text{b}(\bm n, \{ \bar{\bm q}\})/k_B T\right]} ,
\end{align}
where $\nu(\bm n, \{ \bar{\bm q} \})$ is a measure of the jump-attempt frequency and $E_\text{b}(\bm n, \{ \bar{\bm q} \})$ is the energy barrier for the atomic hop. Both $\nu$ and $E_\text{b}$ are functions of the configuration vectors $\bm n_p$ and mean positions $\bar{\bm q}_p$ of the neighboring sites as well as of the chemical species $\alpha$ hopping from site $i$ to $j$. Therefore, using the expression above for a species $\alpha$ would require computing the jump frequency $\nu$ and energy barrier $E_\text{b}$ for a prohibitively large number ($\approx (M+1)^{N}$ for $M$ chemical species) of combinatorial possibilities corresponding to the neighboring configurational vectors $\bm n_p$. To avoid this, we replace averaging over these combinatorial possibilities by a single NEB computation with the neighboring sites being described as concentration-averaged atoms based on the atomic concentrations at every site at the current time step. In this case, the average transition rate can be approximated as
\begin{align}
    \langle \gamma_{i \rightarrow j}^{\alpha} \rangle \approx \nu(\bm c, \bar{\bm q}) \exp{\left[-E_\text{b}^{i \rightarrow j}(\bm c, \bar{\bm q})/k_B T\right]},
\end{align}
where $\bm c$ and $\bar{\bm q}$ are, respectively, the concentration vectors and mean positions of the atoms which are neighbors of either $i$ or $j$. It is important to note that the attempt frequency and energy barrier still depend on the type of atom (species~$\alpha$) that attempts the hop. Therefore, distinct NEB computations are still required to quantify the energy barriers corresponding to hops of different atomic species in the same environment.

According to H-TST, the jump-attempt frequency is the ratio of the vibrational eigenfrequencies of the entire system in the minimum corresponding to the atom being at site $i$ to the vibrational eigenfrequencies for a system constrained on the saddle plane dividing the two minima~\citep{vineyard1957frequency}. However, such a calculation is prohibitively expensive and the attempt frequency $\nu( \bm c, \bar{\bm q} )$ is often approximated as the Einstein frequency of atom~$i$~\citep{vineyard1957frequency}. It is interesting to note that the vibrational frequency of an atom can be mapped to the thermodynamically relaxed positional variance $\Sigma$ in the GPP model. Substituting $\Sigma_i = k_B T / m \omega_i^2 $ in Eq.~\eqref{gpp matrix details} recovers the DMD \cite{li2011diffusive} and max-ent~\cite{kulkarni2008variational} frameworks, where $\omega_i$ is the Einstein frequency for the vibration of atom $i$ about its mean position. In the present GPP framework, $\Sigma_i$ is a free variable to be determined from Eq.~\eqref{eq: EOM QS} for every atomic site, so the approximation $\nu\approx \omega_i/2\pi$ leads to the average transition rate being computed as
\begin{align}
        \langle \gamma_{i \rightarrow j}^{\alpha} \rangle = \frac{1}{2 \pi} \sqrt{ \frac{k_B T}{m_{\alpha} \Sigma_i } } \exp{\left[-E_\text{b}^{i \rightarrow j}(\bm c, \bar{\bm q})/k_B T\right]}.
    \label{eq: avg rate final}
\end{align}

To compute the energy of the system at the saddle point between the crossover of atom $\alpha$ from site $i$ to $j$, we use the NEB technique with those participating atoms that lie within a cutoff radius $r_\text{NEB}$ from both sites $i$ and~$j$. Multiple replicas of these participating atoms are initialized and connected with string forces. A choice of seven replicas was deemed an ideal compromise between accuracy and efficiency after numerical experimentation. These atoms are surrounded by their neighbors, whose positions stay the same for every replica. They are hence referred to as non-participating atoms in the NEB computation. Before relaxing the entire band of replicas, we find the start and end points of the band. For the starting point, we create a vacancy at site $j$ and thermally relax the ensemble by allowing only the participating atoms to move. For the endpoint, the procedure is repeated after creating a vacancy at $i$ instead. This has two advantages: (i)~obtaining better estimates of the jump frequency of the moving atom ($ \tilde{\Sigma}^\alpha_i $ and $ \tilde{\Sigma}^\alpha_j $) and (ii)~ensuring a robust relaxation of the entire band. Having identified the start and end points, the band of replicas of the participating atoms is relaxed under the force field exerted by the neighboring non-participating atoms, using the Fast Inertial Relaxation Engine (FIRE) solver \citep{PhysRevLett.97.170201}. Force nudging is adopted based on \citet{henkelman2000improved}, and perpendicular band stiffness between band segments is added to avoid kink formation. As we show in Sect.~\ref{subsec: selfD}, a choice of $r_\text{NEB}$ equal to the nearest-neighbor distance is sufficient to capture the diffusion kinetics accurately. Hence, all the nearest-neighbor sites of both $i$ and $j$ are allowed to move when the atom hops from site $i$ to $j$. Once the NEB computation is converged, the total energy of the participating and non-participating atoms is computed for the first, last, and saddle point replicas and inserted into Eq.~\eqref{eq: avg rate final} to obtain the forward and backward transition rates.

\subsection{Finding vacancy segregation and binding energies} \label{learning from Oct24}
As explained in Sect.~\ref{Diffusion in Intermetallics}, computing local vacancy interaction energies and segregation energies is required to obtain the \textit{vacancy enrichment factors} $(\kappa)$ and \textit{vacancy segregation factors} $(\lambda)$ for intermetallic diffusion. 
This section explains the methodology to obtain values of $\kappa^\alpha_{ij}$ and $\lambda^v_j$ for a generic pair of sites $i$ and $j$. 

\begin{figure}[h]
\includegraphics[width=\linewidth]{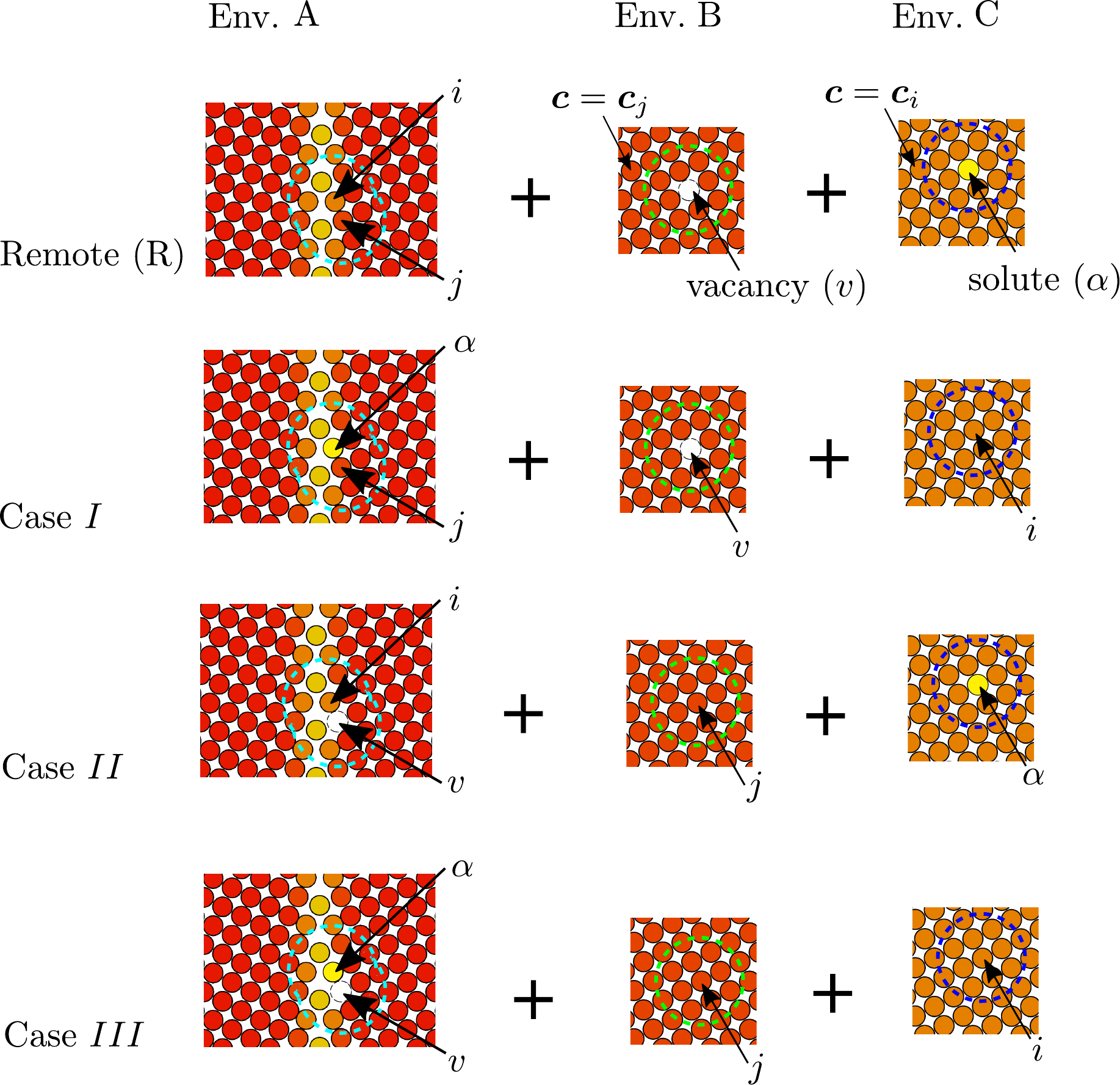}
\caption{\label{fig:binding_energy_illustration} Computation of the environment-dependent vacancy binding energy and segregation energies.}
\end{figure}

We consider three local environments (illustrated in Fig.~\ref{fig:binding_energy_illustration}): (A)~around sites $i$ and $j$, (B)~around a site embedded in a bulk of identical atoms with average concentration $\bm c_j$, and (C)~around a site embedded in a bulk of identical atoms with average concentration $\bm c_i$. Four distinct cases are considered regarding the combinatorial placement of a vacancy and chemical species $\alpha$. The total free energy of each scenario is equivalent to the sum of the free energy of the three local environments (A,B, and C). The remote case (R) corresponds to the scenario of sites $i$ and $j$ being occupied by the hybrid atoms with average concentration $\bm c_i$ and $\bm c_j$, the vacancy $(v)$ is far away in bulk environment~B, and the solute $(\alpha)$ is far away in bulk environment~C. Case~$\uppercase\expandafter{\romannumeral 1}$ corresponds to a swap of solute $\alpha$ with the hybrid atom at site $i$ in environment~A, which is moved to the bulk environment~C with identical neighbors around it. The resulting configuration has an excess free energy $F_{\text{seg}_i}^\alpha$ required to bring the solute atom~$\alpha$ to site~$i$. Similarly, Case~$\uppercase\expandafter{\romannumeral 2}$ corresponds to the swap of vacancy $v$ with the hybrid atom at site $j$ in environment~A, and the excess free energy $F_{\text{seg}_j}^v$ is required to bring the vacancy to site $j$. Case~$\uppercase\expandafter{\romannumeral 3}$ corresponds to both of these swaps happening simultaneously. The resulting configuration now has an excess free energy equal to the sum of the aforementioned segregation energies and also an environment-dependent interaction free energy between the solute and vacancy, denoted by $F_{\text{int}_{ij}}^{\alpha v}$. The total free energy in each of the four scenarios can be written as
\begin{subequations}
\begin{equation}
    F_R = F_R^\text{A} + F_R^\text{B} + F_R^\text{C},
\end{equation}
\begin{equation}
    F_{\uppercase\expandafter{\romannumeral 1}} =
F_{\uppercase\expandafter{\romannumeral 1}}^\text{A} + F_{\uppercase\expandafter{\romannumeral 1}}^\text{B} +
F_{\uppercase\expandafter{\romannumeral 1}}^\text{C} =
F_{\text{R}} + F_{\text{seg}_i}^\alpha,
\end{equation}
\begin{equation}
F_{\uppercase\expandafter{\romannumeral 2}} =
F_{\uppercase\expandafter{\romannumeral 2}}^\text{A} + F_{\uppercase\expandafter{\romannumeral 2}}^\text{B} +
F_{\uppercase\expandafter{\romannumeral 2}}^\text{C} =
F_{\text{R}} + F_{\text{seg}_j}^v,
\end{equation}
\begin{eqnarray}
F_{\uppercase\expandafter{\romannumeral 3}} =
F_{\uppercase\expandafter{\romannumeral 3}}^\text{A} + F_{\uppercase\expandafter{\romannumeral 3}}^\text{B} +
F_{\uppercase\expandafter{\romannumeral 3}}^\text{C}=\nonumber\\
F_{\text{R}} +  F_{\text{seg}_i}^\alpha + F_{\text{seg}_j}^v + F_{\text{int}_{ij}}^{\alpha v}.
\end{eqnarray}
\end{subequations}

Once a reasonable estimate of the free energies $(F_R,F_{\uppercase\expandafter{\romannumeral 1}},F_{\uppercase\expandafter{\romannumeral 2}},F_{\uppercase\expandafter{\romannumeral 3}})$ has been obtained from simulations, the above set of algebraic equations can be solved simultaneously for the local environment-dependent vacancy segregation and interaction energies, which in turn yields the vacancy enrichment factor $\kappa_{ij}^\alpha$ and the vacancy segregation factor $\lambda^v_j$. The energy of the bulk environments~B and C can be computed offline and stored in a lookup table for a range of atomic concentrations of species~$\alpha$. 
The energy of environment~A in the remote case ($F_R^\text{A}$) is already available as part of the simulation, and $F_{\uppercase\expandafter{\romannumeral 3}}^\text{A}$ is available as the energy of the first replica of the NEB computation performed to obtain the transition energy barrier for the movement of species $\alpha$ from site $i$ to $j$. $F_{\uppercase\expandafter{\romannumeral 1}}^\text{A}$ and $F_{\uppercase\expandafter{\romannumeral 2}}^\text{A}$ must be additionally computed on-the-fly. To obtain these, we again relax a small neighborhood of participating atoms (within a chosen cutoff) in a force field generated by the surrounding non-participating atoms as explained in Sect.~\ref{subsec: NEB impl.}.

\subsection{ Non-dimensionalization and Numerical Stability }
The absolute value of transition rates $\langle \gamma_{i \rightarrow j}^{\alpha} \rangle$ in solids at temperatures of engineering interest is usually very low. Therefore, a non-dimensionalization of those terms in Eq.~\eqref{eq: avg rate final} renders numerical time integration of Eq.~\eqref{eq: master_eq_simplified} feasible. To this end, we introduce a reference mass $m_0$, reference covariance $\Sigma_0$, and reference energy barrier $E_{\text{b}_0}$ to define $m_{\alpha} = m_0 \hat{m}_{\alpha} $,  $\Sigma_i = \Sigma_0 \hat{\Sigma}_i$, and $E_\text{b} = E_{\text{b}_0} + \Delta E_\text{b}$. This turns Eq.~\eqref{eq: avg rate final} into $\langle \gamma_{i \rightarrow j}^{\alpha} \rangle = \gamma_0 \langle \hat{\gamma}_{i \rightarrow j}^{\alpha} \rangle$ where
\begin{align}
&\gamma_0 = \frac{1}{2 \pi} \sqrt{ \frac{k_B T}{m_0 \Sigma_0 } } \exp{\left[-E_{\text{b}_0}/k_B T\right]} \quad \text{and} \nonumber \\
 & \langle \hat{\gamma}_{i \rightarrow j}^{\alpha} \rangle =   \frac{1}{ \sqrt{\hat{m}_{\alpha} \hat{\Sigma}_i } } \exp{\left[- \Delta E_\text{b}(\bm c, \bar{\bm q})/k_B T\right]}.
 \label{eq: NDsation}
\end{align}
The value of $\Sigma_0$ is chosen to be the position variance of a bulk atom at thermal equilibrium at temperature~$T$. A reference energy barrier $E_{\text{b}_0}$ is chosen close to the energy of the transition of the solute or solvent to its nearest neighbor in a bulk neighborhood. Substituting this in Eq.~\eqref{eq: master_eq_simplified} suggests a dimensionless time $\hat{t}^v = \gamma_0 t $ such that
    \begin{align}
    \frac{\dd c_i^\alpha }{\dd \hat{t}^v} = \sum_{j \in NN^i} \left [ c_i^v c_j^\alpha \langle \hat{\gamma}_{j \rightarrow i}^{\alpha} \rangle - c_j^v c_i^\alpha \langle \hat{\gamma}_{i \rightarrow j}^{\alpha}\rangle    \right ],
    \label{eq: master eq we use ND}
\end{align}
where the superscript $v$ indicates that this dimensionless master equation will be used to model the vacancy diffusion kinetics in an otherwise pure sample. 

For intermetallics, we use Eq.~\eqref{eq: approx master eq we use alloys} for concentration updates. Substituting the non-dimensionalization discussed above in Eq.~\eqref{eq: approx master eq we use alloys}, one obtains
\begin{align}
    \frac{\dd c_i^\alpha }{\dd \hat{t}} = \sum_{j \in NN^i} \left [ \kappa^\alpha_{ji} \lambda_i^v c_j^\alpha \langle \hat{\gamma}_{j \rightarrow i}^{\alpha} \rangle 
    -  \kappa^\alpha_{ij} \lambda_j^v c_i^\alpha \langle\hat{\gamma}_{i \rightarrow j}^{\alpha}\rangle    \right ]
    \label{eq: approx master eq we use alloys ND}
\end{align}
with the dimensionless time $\hat{t} = c_0^v \gamma_0 t$. 

We use an explicit Euler time integration scheme to integrate Eq.~\eqref{eq: master eq we use ND} for pure metals and Eq.~\eqref{eq: approx master eq we use alloys ND} for alloys in time. Therefore, time steps $\Delta \hat{t}^v$ and $\Delta \hat{t}$ must be chosen below an upper bound stemming from the stability constraint of the explicit integration scheme. Von Neumann analysis yields the upper limit for the time step, which we endow with a safety factor of $\delta$ in our simulations, resulting in the upper limits for an FCC lattice as
\begin{align}
\Delta \hat{t}_{\max}^v &= \frac{\delta}{12\, \text{max}(c^v \langle \hat{\gamma^\alpha} \rangle )}, \nonumber \\
    \Delta \hat{t}_{\max} &= \frac{\delta}{12\, \text{max}(\kappa^{\alpha} \lambda^v \langle \hat{\gamma^\alpha} \rangle )}
\end{align}
for Eqs.~\eqref{eq: master eq we use ND} and~\eqref{eq: approx master eq we use alloys ND}, respectively, where $\text{max}(\cdot)$ is the maximum value of the argument for any two nearest neighbors in the entire atomic ensemble. The value of $\delta$ used in simulations is stated for the all cases in Sect.~\ref{sec: examples}.

\subsection{ Numerical implementation details  }
Computing the forward and backward average transition rates $\langle \gamma_{i \rightarrow j}^{\alpha} \rangle$ and $\langle \gamma_{j \rightarrow i}^{\alpha} \rangle$, respectively, is required for all nearest-neighbor pairs in the atomic ensemble and all solute species $\alpha$ being considered. This can render a single concentration update step too expensive. However, it is important to realize that many of these pairs of $i,j$ nearest neighbors have an identical neighborhood with respect to translations and proper rotations. Consequently, the activation energy of atomic hops is invariant to translations and rotations. Therefore, many NEB computations are redundant, and their computational cost can be avoided by identifying some invariant measures of the local environment based on the mean positions $\bar{\bm q}$ of the participating and non-participating atoms of the NEB computation. This is analogous to finding an optimum set of descriptors for the local atomic environment \citep{PhysRevB.87.184115, PhysRevLett.98.146401}, which is a crucial component for building machine-learned interatomic potentials \citep{doi:10.1021/acs.jpca.9b08723, PhysRevB.100.024112}. 

In this work, we use the number of neighbors and the centrosymmetry of sites $i$ and $j$ as the descriptors, because they are sufficient to identify unique environments for the applications we consider in Sect.~\ref{sec: examples}. At every concentration update step, unique $i,j$ neighbor pairs are detected by using those sets of descriptors, and NEB computations are performed for them. The resulting activation barriers, vacancy segregation and binding energies are tabulated along with the descriptors for the corresponding pair. Finally, this information of every pair $(i,j)$ in the ensemble is obtained by matching their descriptors with one of the unique sets of descriptors in the lookup table.

To make the entire procedure computationally feasible, energy barriers, vacancy segregation, and binding energies are not computed at every concentration update time step. Instead, an adaptive rate monitoring algorithm is used, where this pair-wise information is computed every $n$ time steps, and $n$ is increased/decreased by a factor $k$ if the globally maximum concentration update rate decreases/increases by a factor $l$, respectively. The values of $\{n,k,l\}$ for this algorithm are user-input parameters and have been tabulated for the simulations in Sect.~\ref{sec: examples}. 

Finally, the intermediate thermal and structural atomic relaxations, solving Eqs.~\eqref{eq: EOM QS}, are triggered only when the L$_2$-norm of the force increases by a factor $g$ compared to the one of the previous relaxation. $g=100$ has been chosen for all simulations in this work.

\section{Results and Discussion} \label{sec: examples}
To demonstrate the performance of the framework introduced above, we discuss the results of predicting the self-diffusion coefficient in bulk copper and simulating long-term coupled chemo-thermo-mechanics for self-diffusion near a stacking fault in aluminum (Sect.~\ref{subsec: selfD}), followed by the kinetics and equilibrium values of magnesium segregation to a stacking fault and a symmetric tilt grain boundary in Al-Mg alloys (Sect.~\ref{subsec: alloyD}).

\subsection{Self Diffusion} \label{subsec: selfD}

\subsubsection{ Bulk diffusion in Cu }
We start by validating the expression for the transition rate in Eq.~\eqref{eq: avg rate final} by comparing the self-diffusion coefficient in bulk copper, using the EAM potential developed by \citet{mishin2001structural}, to experimental studies summarized in \cite{mehrer2007diffusion}. While the experimental data is limited to relatively high temperature levels, we compute the self-diffusion constants for a wide range of temperatures, ranging from 200~K to 1200~K.  The diffusion constant can be inferred from the master equation by taking a long wavelength limit of Eq.~\eqref{eq: master_eq_simplified} and considering only vacancies as impurities. In this case, the mass transport kinetics is governed by Eq.~\eqref{eq: master_eq_simplified} with $\alpha$ being the solvent atom (here considered to be copper). For a perfect bulk crystal, the vacancy concentration at all sites is obtained by computing the vacancy formation free energy at that temperature $(F_\text{vf})$ and can be written as $c^v = \exp{\left(-F_\text{vf}/k_BT\right)}$ \citep{mehrer2007diffusion}. Moreover, the transition rates for the forward and backward jumps are identical, as every site is equivalent in a perfect bulk (i.e., $\langle \gamma_{i \rightarrow j}^\text{Cu} \rangle = \langle \gamma_{j \rightarrow i}^\text{Cu} \rangle = \gamma^\text{Cu}$). Finally, the solvent concentration is assumed to vary slowly in the bulk. Hence, the solvent concentration at any nearest-neighbor site $j$ can be written as
\begin{align}
    c^\text{Cu}_j=c^\text{Cu}_i+\left.\nabla c^\text{Cu}\right|_i \cdot \bm r_{j i}+\left.\frac{1}{2} \nabla \nabla c^\text{Cu}\right|_i:\left(\bm r_{j i} \otimes \bm r_{j i}\right),
    \label{eq: long wave limit}
\end{align}
where $\bm r_{ji}$ is the vector pointing from atom $i$ to $j$.
Substituting the above assumptions in Eq.~\eqref{eq: master_eq_simplified} leads to
\begin{align}
    \dot{c}^\text{Cu}_i= \left[ \frac{c^v \gamma^\text{Cu}}{2}\left(\sum_{N N} \bm r_{j i} \otimes \bm r_{j i}\right) \right] :\left.\nabla \nabla c^\text{Cu}\right|_i.
\end{align}
When comparing the above to Fick's law of mass diffusion, the quantity in square brackets can be identified as the diffusivity tensor $\bm D$. After inserting the expressions for $c^v$ and $\gamma^\text{Cu}$, the trace of this tensor can be simplified for an FCC lattice to
\begin{align}
    D = 
    \frac{a^2}{2 \pi} \sqrt{ \frac{k_B T}{m_\text{Cu} \Sigma_i^\text{Cu} } } \exp{\left[-(E^\text{Cu}_\text{b}+F_\text{vf})/k_B T\right]},
\end{align}
where $a$ is the relaxed lattice constant at the temperature of interest. For every temperature, we consider a fully periodic $8 \times8\times8$ supercell of copper in the FCC lattice structure. This supercell (containing $N$ atomic sites) is first relaxed (using Eqs.~\eqref{eq: EOM QS}) at that temperature to obtain the lattice spacing $a$, position variance $\Sigma$, and free energy $F_{\mathrm{bulk}}$ at thermal equilibrium. Then, the vacancy concentration at the center of the supercell is set to $1$, and the structure is relaxed again without allowing the box dimensions to relax. Constraining periodic relaxation here is important to simulate the effect of creating a single vacancy in an infinite bulk lattice. The thus-obtained free energy is denoted as $F_{vac}$. The free energy associated with vacancy formation is then computed as
\begin{align}
    F_\text{vf} = F_\text{vac} - \frac{N-1}{N} F_\text{bulk}.
\end{align}

\begin{figure}[h!]
\includegraphics[width=\linewidth]{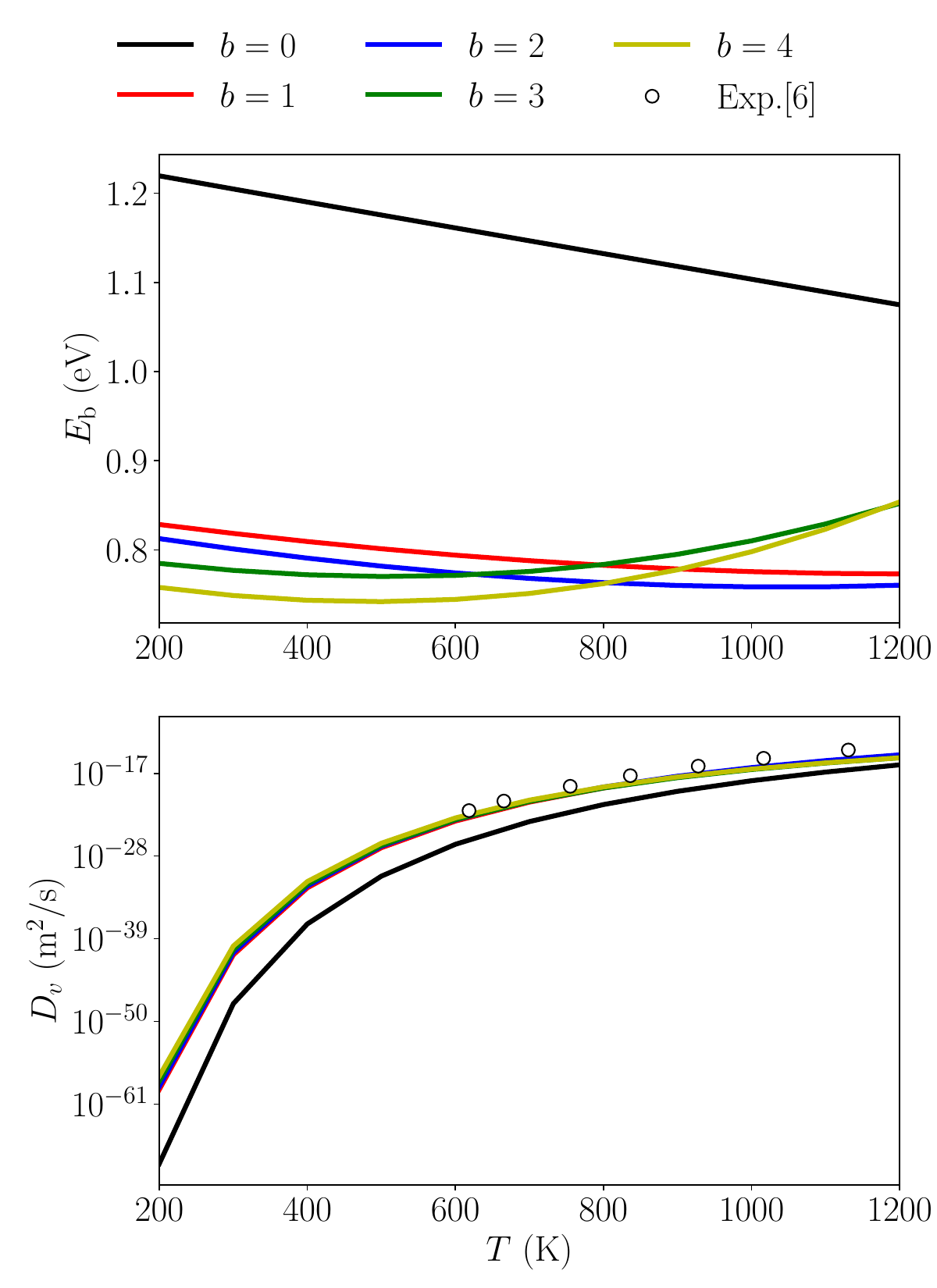}
\put(-230,160){$(a)$}
\put(-230,0){$(b)$}
\caption{\label{fig:Cu bulk diffusion} (a) Energy barrier for a nearest-neighbor hop and (b) self-diffusion coefficient in bulk copper as a function of temperature as obtained by moving different layers of neighboring atoms, and its comparison to experimental values from \citet{mehrer2007diffusion}. }
\end{figure}

Next, the pure, thermally relaxed copper supercell is used to compute the transition energy barrier and jump frequency for an atom at the center of the cell to one of its nearest-neighbor sites. To this end, we use the implementation discussed in Sect.~\ref{subsec: NEB impl.} with different values of the cutoff radius to identify participating atoms. The cutoff radii are chosen such that a union of $b\in \{0,1,2,3,4\}$ nearest-neighbor shells of sites $i$ and $j$ are allowed to move in the band relaxation. The resulting energy barrier and diffusion constants are shown in Fig.~\ref{fig:Cu bulk diffusion}. Not including any neighbors of the moving atom in the NEB calculation results in an excessively high energy barrier, which in turn results in a diffusion constant that is orders of magnitude below that observed in experiments. When first- or more nearest neighbors to the moving atom are allowed to participate in the NEB relaxation, the diffusion constant matches well with the experimental data from \cite{mehrer2007diffusion}. Because including only the first nearest neighbors already results in good agreement of bulk diffusion constant with experimental data, we choose this as the default for the simulations in Sect.~\ref{subsec: alloyD}.

\subsubsection{ Stacking fault in Al }

\begin{figure}[htbp]
\includegraphics[width=\linewidth]{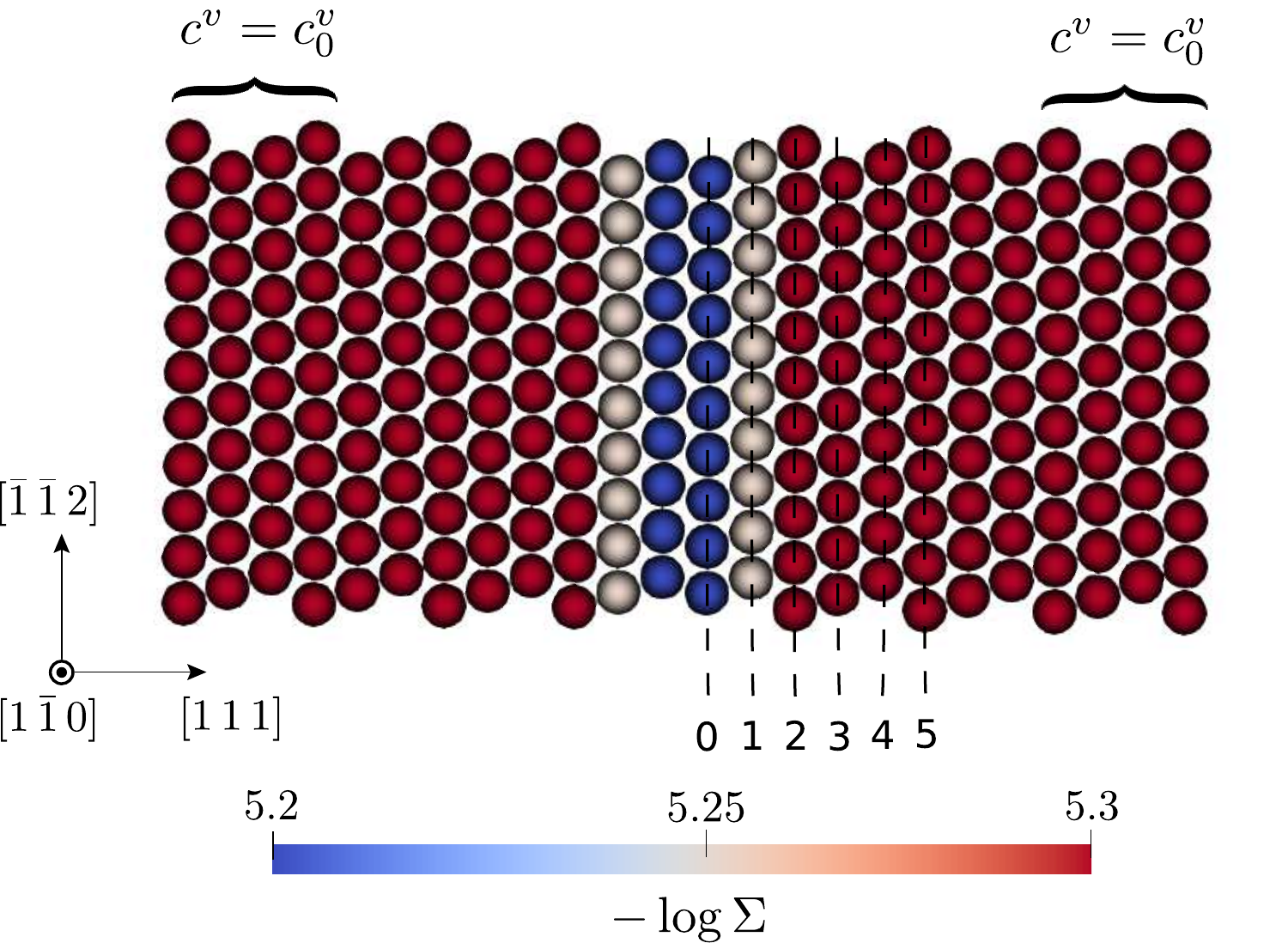}
\caption{\label{fig:SF geom} Stacking fault configuration, color-coded by the relaxed atomic position variances $\Sigma_i$ at $300$~K, indicating the $0^\text{th}$ to $5^\text{th}$ layer of atoms next to the stacking fault. }
\end{figure}

As a further benchmark, we study the vacancy diffusion kinetics and equilibrium profile near a stacking fault in aluminum, using the EAM potential developed by \citet{liu1998grain} at temperatures of 300, 400, 500, and 600~K. This scenario is important, as the concentration of vacancies near a stacking fault can alter the generalized stacking fault energy (GSFE) and hence the response to mechanical loading \citep{asadi2014effect}.
We start by creating a rotated FCC supercell with the lattice parameter $a$ of pure aluminum (tabulated in Tab.~\ref{tab:vac seg kinetics}), relaxed at the temperature of interest using GPP, such that the Cartesian $x$-axis aligns with the $[111]$ crystallographic plane. We construct a stacking fault in the otherwise perfect FCC supercell by removing a plane of atoms along the $x$-axis and shifting all planes above by the interplanar distance $a/\sqrt{3}$ along the negative $x$-axis. The resulting configuration is relaxed within the GPP framework, after initializing all atomic sites with a vacancy concentration of $c_0^v = 0.01\%$. The final configuration is shown in Fig.~\ref{fig:SF geom}.

\begin{figure}[htbp]
\includegraphics[width=\linewidth]{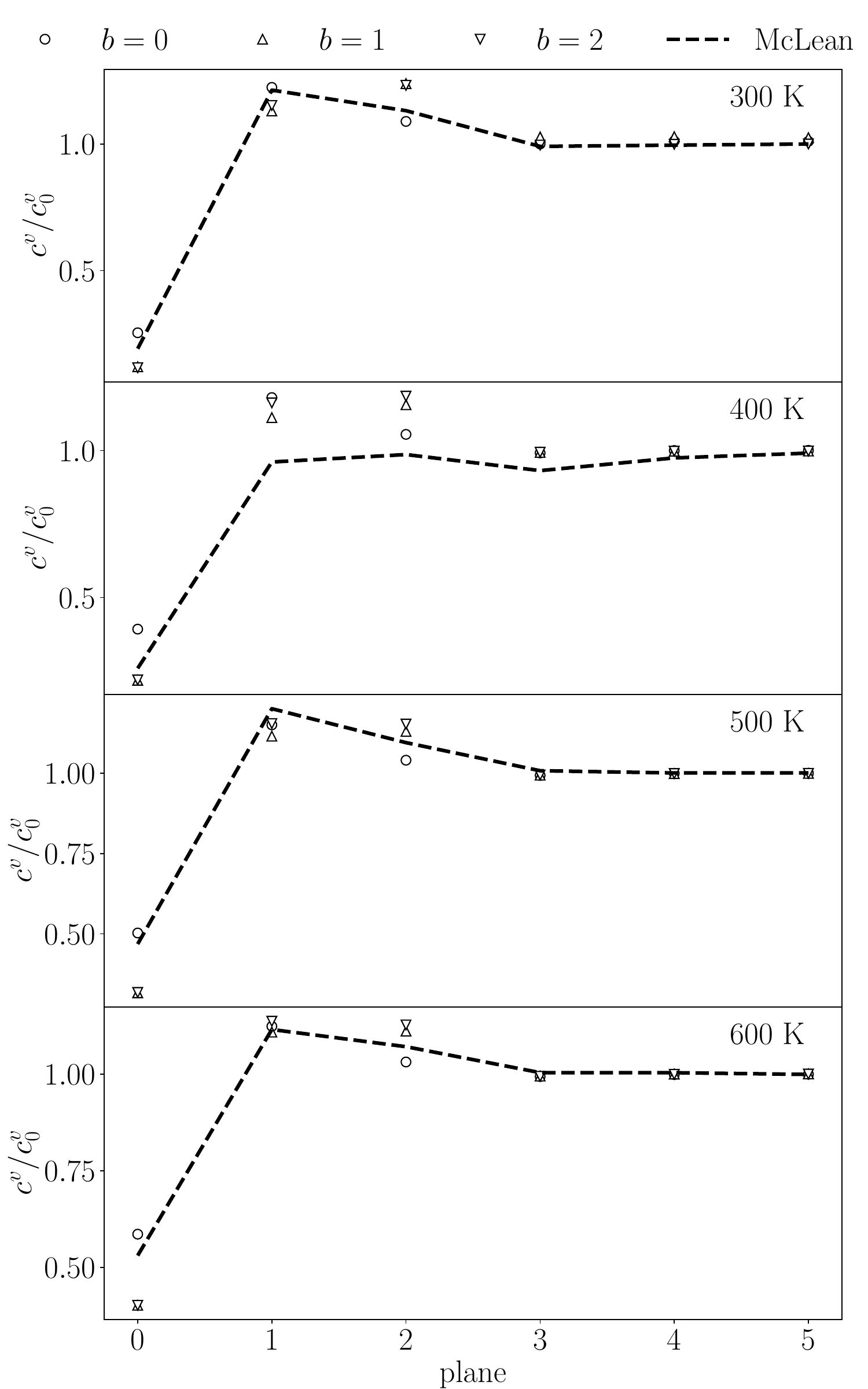}
\caption{\label{fig:SF vacancy equilibirum} Equilibrium vacancy concentration profiles near the stacking fault for different numbers of participating nearest-neighbor shells at different temperatures, compared to Langmuir-McLean isotherms. }
\end{figure}

\begin{figure}[htbp]
\includegraphics[width=\linewidth]{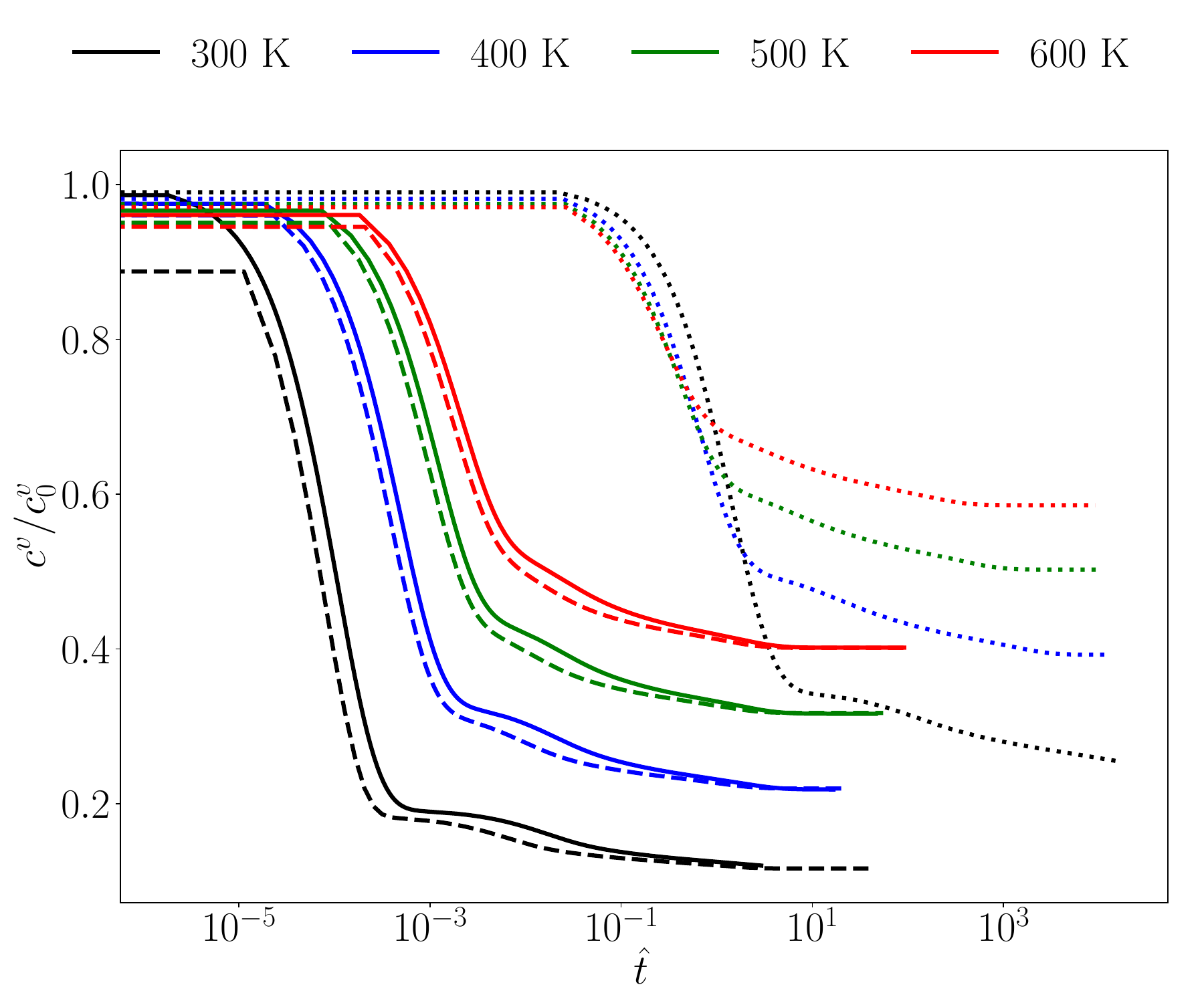}
\caption{\label{fig:SF kinetics vac} Segregation kinetics for vacancies at the stacking fault plane for $b=0$ (dotted lines), $b=1$ (solid lines), and $b=2$ (dashed lines) at different temperatures. }
\end{figure}

The resulting atomic ensemble is subjected to a staggered scheme of concentration update (using Eq.~\eqref{eq: master eq we use ND}) and thermal relaxation (using Eq.~\eqref{eq: EOM QS}) to track the vacancy equilibration kinetics, using a fixed concentration boundary condition on planes farther than 8 layers away from the stacking fault, as shown in Fig.~\ref{fig:SF geom}. This was performed for three choices of the cutoff radius to select participating atoms in the NEB computations, such that a union of $b \in \{0,1,2\}$ nearest-neighbor shells of sites $i$ and $j$ are allowed to move in the band relaxation. The resulting equilibrium vacancy concentration profiles at all studied temperatures are shown in Fig.~\ref{fig:SF vacancy equilibirum} along with the concentration profile obtained from the dilute limit (Langmuir-McLean isotherm solution). The latter is obtained at plane $p$ from the stacking fault as 
\begin{equation}
    \frac{c^v(p)}{c^v_0} = \frac{1}{c^v_0 + (1-c^v_0)\exp{\left(F(p)/k_BT\right)}},
\end{equation}
where $F(p)$ is the segregation energy of the impurity (vacancy in this case) at plane~$p$, which is obtained by replacing an atom with a vacancy at plane $p$ in an otherwise pure crystal of aluminum and relaxing this atomic ensemble, using Eqs.~\eqref{eq: EOM QS}.
The diffusion kinetics for plane~0 are shown in Fig.~\ref{fig:SF kinetics vac} for all simulated temperatures. Note that the concentration evolution is presented as a function of the dimensionless time $\hat{t}^v = t\times  \gamma_0$. The reference values of the position variance ($\Sigma_0$) and energy barrier ($E_{\text{b}_0}$) are summarized in Tab.~\ref{tab:vac seg kinetics} along with the (actual) time $t_1$ taken by the stacking fault plane to reach within $1 \%$ of its equilibrium concentration value. 

As expected, reduced repulsion of the vacancies away from the stacking fault (and, therefore, a higher enrichment factor) is observed with increasing temperature because the relative free energy gained by the system from configurational rearrangements reduces, hence reducing the thermodynamic driving force for such rearrangements \citep{C7TA01080J}. Additionally, moving no nearest neighbors ($b=0$) slightly overpredicts the segregation factor at the stacking fault compared to the $b=1$ and $b=2$ predictions (which are almost identical). It is interesting to notice that, although the equilibrium values of the enrichment factors are well captured by all three choices of participating atom cutoff radius, moving none of the nearest neighbors ($b=0$) severely overpredicts the time required to attain equilibrium by several orders of magnitude. By contrast, moving 1 vs.\ 2 nearest neighbors has seemingly negligible effect on the segregation kinetics, as displayed by Fig.~\ref{fig:SF kinetics vac}. This reinforces the previous observation from the bulk self-diffusion in copper, viz.\ that moving nearest neighbors during NEB calculations is important and sufficient to capture the mass diffusion kinetics.

\begin{table}[htb]
    \centering
    \caption{Kinetics of vacancy segregation to the stacking fault plane with a bulk vacancy concentration $c_0^v = 0.01\%$. $t_1^{(\cdot)}$ refers to the time required for the stacking fault plane to reach within $1 \%$ of its equilibirium vacancy concentration for different numbers of participating neighbor shells in the NEB calculations.}
    \begin{tabular}{ccccc}
        \toprule
         & 300~K & 400~K & 500~K & 600~K \\
        \midrule
        $a $[\AA] & 4.055 & 4.061 & 4.067 & 4.073 \\ 
        $\log \Sigma_0$& -5.27 & -5.00 & -4.78 & -4.60 \\
        $E_{\text{b}_0}\, [eV]$ & 0.85 & 0.85 & 0.85 & 0.85 \\
        $t_{1}^{b=0}$[s] & $3.27\times 10^5$ & $16.6$ & $3.25 \times 10^{-2}$ & $4.89 \times 10^{-4}$ \\
        $t_{1}^{b=1}$[s] & $60.0$ & $2.87\times 10^{-2}$ & $1.88\times 10^{-4}$ & $6.46 \times 10^{-6}$ \\
        $t_{1}^{b=2}$[s] & $56.4$ & $1.52 \times 10^{-2}$ & $1.08 \times 10^{-4}$ & $4.02 \times 10^{-6}$ \\
        \bottomrule
    \end{tabular}
    \label{tab:vac seg kinetics}
\end{table}

\subsection{Diffusion in Alloys} \label{subsec: alloyD}

\subsubsection{ Stacking fault in Al-Mg } \label{subsubsec: SF}
We now consider a more complicated scenario with vacancies and magnesium atoms as impurities. We use Eq.~\eqref{eq: approx master eq we use alloys} with staggered atomic relaxations according to Eq.~\eqref{eq: EOM QS} to obtain long-term segregation kinetics of Mg to the stacking fault for temperatures of 300, 400, 500, and 600~K in an aluminum crystal described by the EAM potential of \citet{liu1998grain}. The chosen input parameters of the adaptive rate-monitoring algorithm to determine the time steps when recomputing the pair-wise vacancy segregation and binding energies and Mg energy barriers are given in Tab.~\ref{tab:ALA parameters}.

\begin{table}[htb]
    \centering
    \caption{ Parameters chosen for the adaptive rate-monitoring algorithm to simulate Mg segregation near a stacking fault in Al. The parameters listed for a concentration are used for all temperatures shown in this work. }
    \begin{tabular}{ccccc}
        \toprule
  $c_0^\text{Mg}$  & $n$ & $k$ & $l$ & $\delta$ \\
        \midrule
        $c_0^\text{Mg} < 0.001$ & 10 & 20 & 10 & 0.9 \\
       $0.001 \leq c_0^\text{Mg} < 0.01 $ & 10 & 10 & 10 & 0.8 \\
       $0.01 \leq c_0^\text{Mg} < 0.1 $ & 4 & 4 & 10 & 0.7 \\
        \bottomrule
    \end{tabular}
    \label{tab:ALA parameters}
\end{table}

\begin{figure}[htbp]
\includegraphics[width=\linewidth]{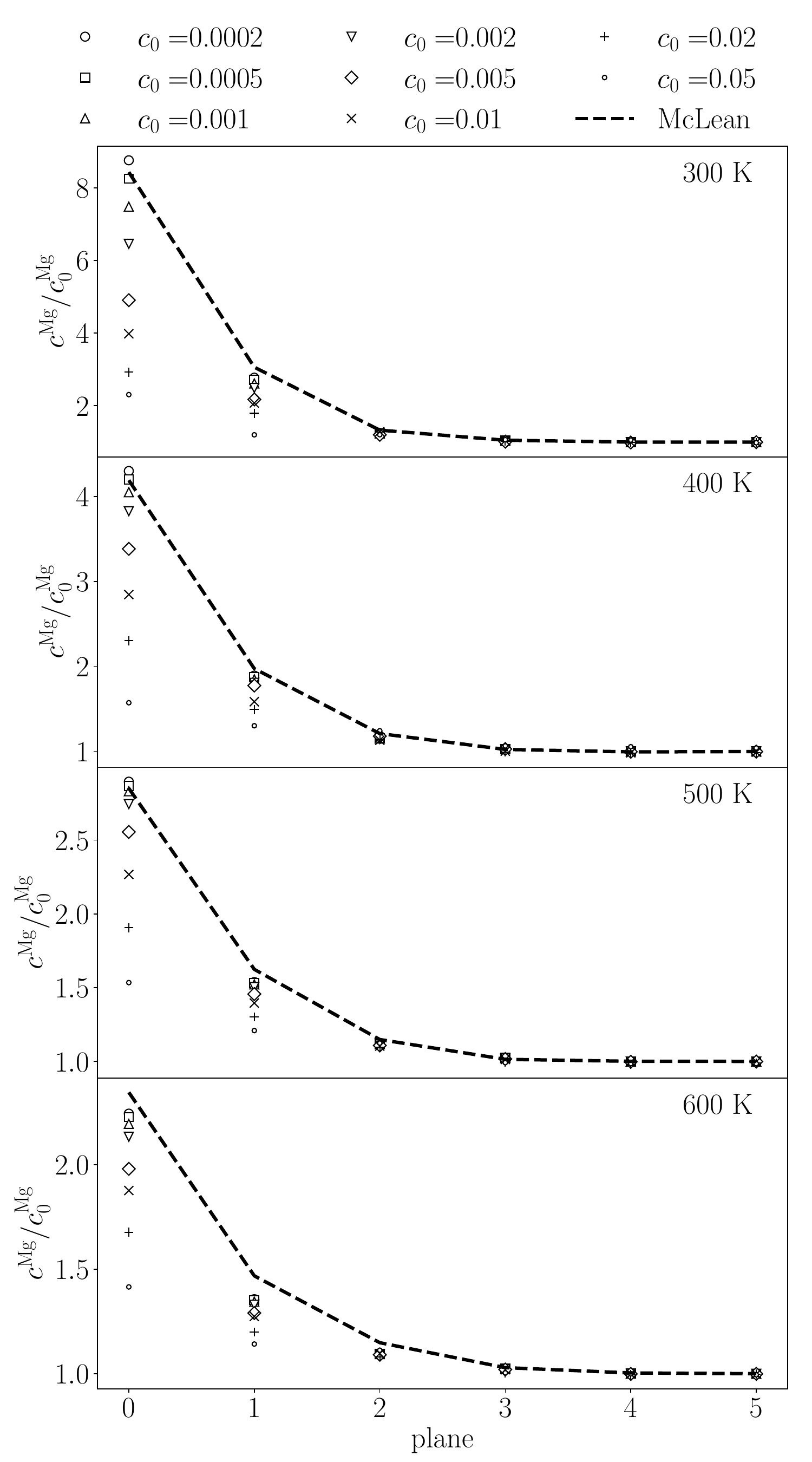}
\caption{\label{fig:SF equilibrium} Equilibrium Mg concentration profiles near the stacking fault for different bulk Mg concentrations at different temperatures, compared to Langmuir-McLean isotherms. }
\end{figure}

Fig.~\ref{fig:SF equilibrium} shows the Mg concentration enrichment near the stacking fault for different bulk Mg concentrations and different temperatures. The concentration profile approaches the Langmuir-McLean isotherm, as the bulk Mg concentration is reduced. This shows the capability of this approach to obtain equilibrium solute concentrations even for non-dilute cases, and its convergence to the well-known McLean solution for a dilute alloy.

\begin{figure}[h!]
\includegraphics[width=\linewidth]{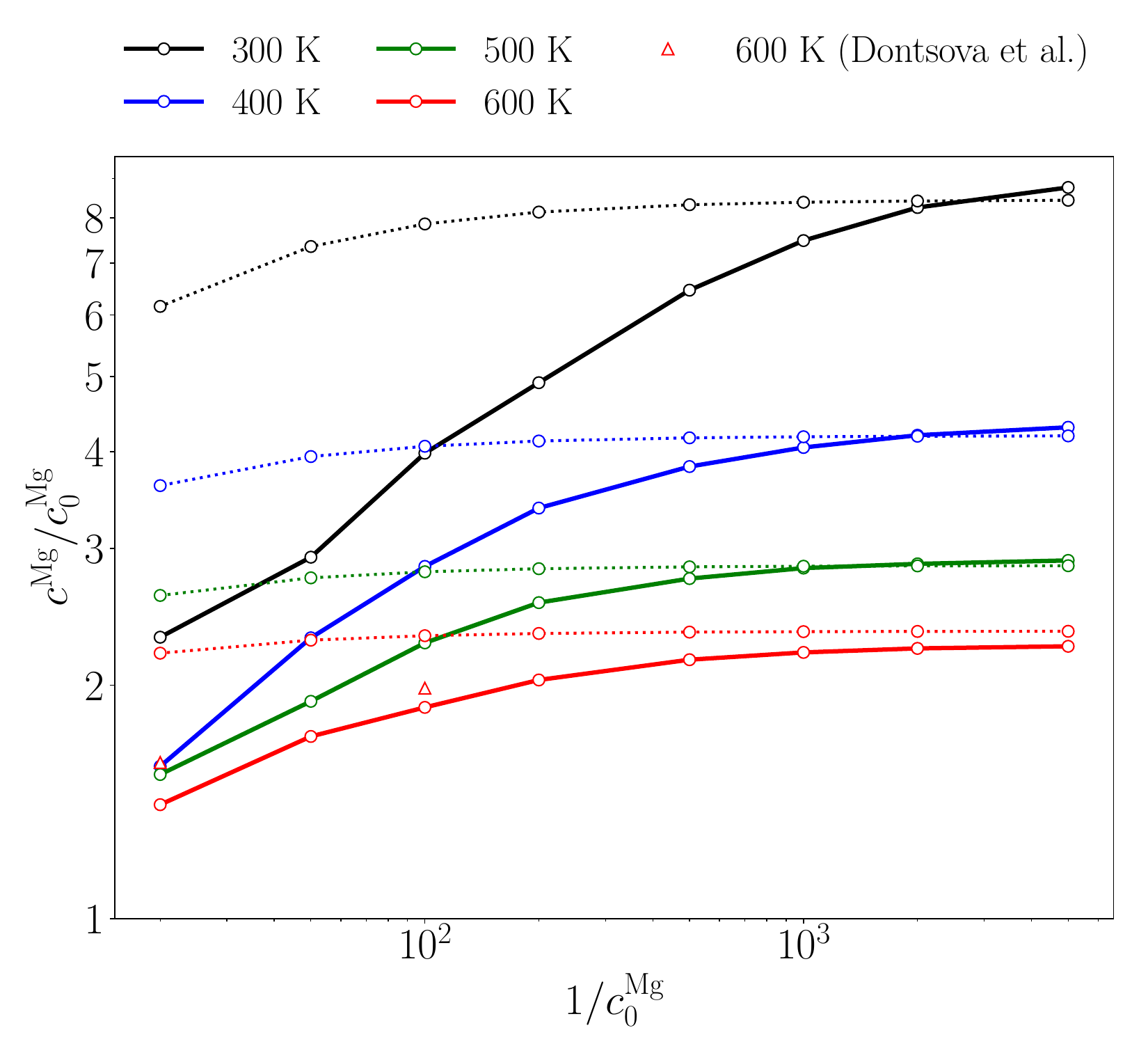}
\caption{\label{fig:SF EF convergence} Convergence of enrichment factors at the stacking fault plane for different temperatures. The McLean solutions (valid only in the dilute limit) are shown as dashed lines for all temperatures. }
\end{figure}

It is also interesting to observe the convergence trend for the Mg enrichment factor at the stacking fault plane as a function of the bulk Mg concentration at different temperatures, which is shown in the log-log plot in Fig.~\ref{fig:SF EF convergence}. The enrichment factors for 600~K at $1\%$ and $5\%$ solid solution of Mg in Al have also been compared to those obtained from an empirical master equation used in DMD by \citet{dontsova2014solute}. Although all plots do converge to the McLean solution at low bulk concentrations, the rate of convergence also depends on the temperature. This highlights the fact that temperature plays an important role in deciding whether or not an alloy is dilute (within the McLean regime).

\begin{figure*}[htbp]
\includegraphics[width=\linewidth]{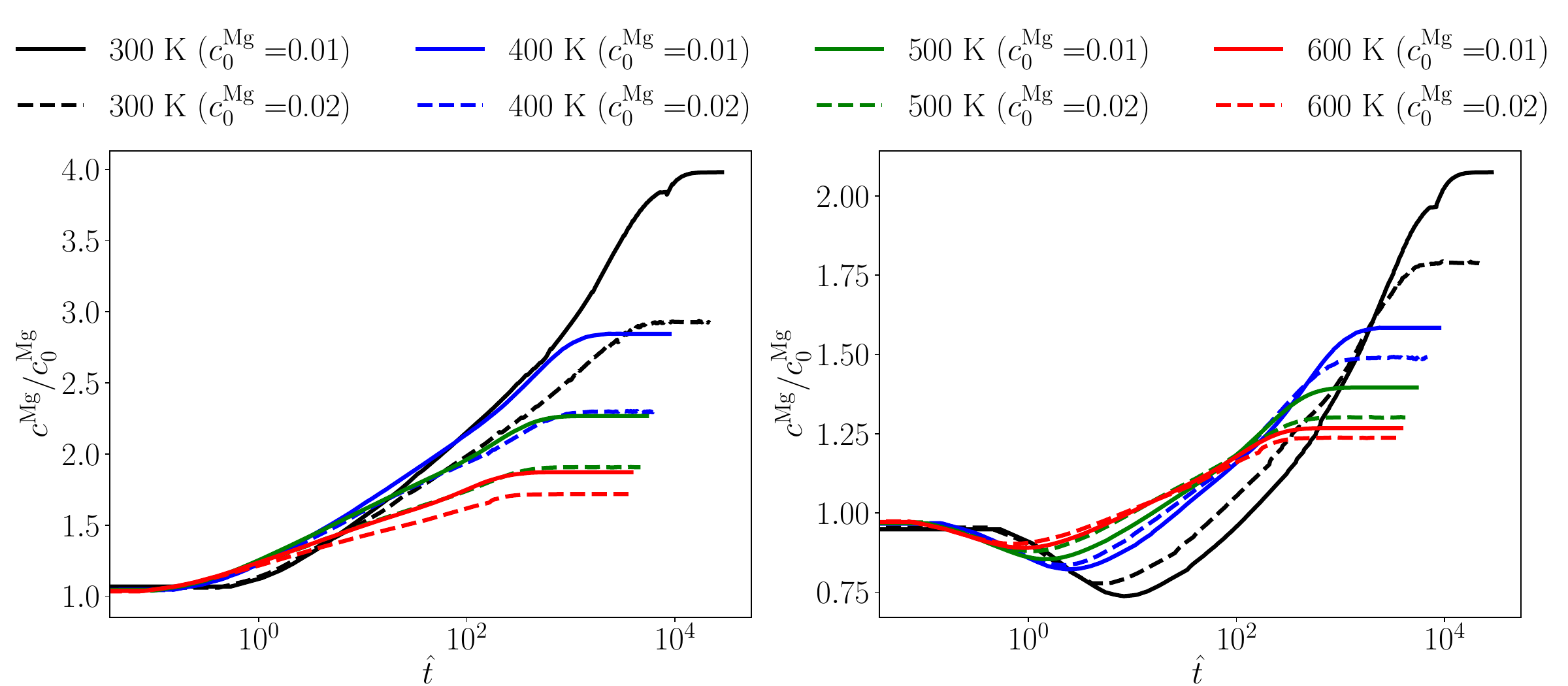}
\put(-500,0){$(a)$}
\put(-245,0){$(b)$}
\caption{\label{fig:SF mg seg kinetics} Magnesium segregation kinetics for different temperatures and bulk magnesium concentrations at (a) plane 0 and (b) plane 1 vs.\ dimensionless time $\hat t$. }
\end{figure*}

We also investigate the segregation kinetics for the plane at and immediately next to the stacking fault for a $1 \%$ and $2 \%$ bulk Mg concentration vs.\ dimensionless time $\hat{t} = c_0^v \gamma_0t$ at different temperatures. Results are shown in Fig.~\ref{fig:SF mg seg kinetics}, showing a gradual transition towards the final equilibrium values at all studied temperatures and bulk concentrations. To convert the dimensionless time to real time, we must estimate the bulk equilibrium vacancy concentration $c^v_0$ at the given temperature and solute concentration. To compute this data for a $1 \%$-Mg and $2 \%$-Mg in Al solid solution, an approach similar to Sect.~\ref{subsec: selfD} is adopted, except that the $8\times8\times8$ supercell now consists of hybrid atoms with $1\%$ Mg and $99\%$ Al (or $2\%$ Mg and $98\%$ Al). The thus-computed free energies of vacancy formation ($F_\text{vf}$) and equilibrium vacancy concentrations ($c_0^v = \exp\left(-F_{\text{vf}}/k_BT\right)$) are included in Tab.~\ref{tab:Mg seg kinetics} for a bulk Mg concentration of $1\%$ and $2\%$, respectively. Further, a reference position variance of $\log \Sigma_0 = -5.545$ and energy barrier $E_{\text{b}_0} = 0.6\, $eV were used to compute the reference transition rate $\gamma_0$ for all simulations in this section. The dimensionless $(\hat{t}_1)$ and actual times $(t_1)$ needed to reach a Mg concentration within $ 1\%$ of its equilibrium value are reported in Tab.~\ref{tab:Mg seg kinetics} for the two bulk Mg concentrations. The equilibration time is slightly reduced for a higher Mg concentration at all temperatures. Further, the estimated time scale required to reach equilibrium at room temperature ($\approx 300$~K) is immensely large ($>10^{5}$ years), whereas it is approximately $2$ seconds for the same alloy at $600$~K. It is important to note that although we used an equilibrium value of $c_0^v$ to estimate these time scales, the dimensionless times required for equilibrium can be scaled up to a real time estimate also for any non-equilibrium bulk vacancy concentration, commonly found in quenched alloys.

\begin{table}[htb]
    \centering
    \caption{Magnesium segregation kinetics to the stacking fault plane for $c_0^\text{Mg} = 0.01$ and $0.02$. $t_1$ refers to the time required for the stacking fault plane to reach within $1 \%$ of its equilibrium Mg concentration.}
    \begin{tabular}{ccccc}
        \toprule
        & 300~K & 400~K & 500~K & 600~K \\
        \midrule

        \multirow{2}{*}{$F_\text{vf}   \, [eV]$} 
        & 0.69 & 0.72 & 0.69 & 0.71 \\
        & 0.69 & 0.69 & 0.69 & 0.67 \\
        \cmidrule(lr){2-5}

        \multirow{2}{*}{$c^v_0$} 
        & $2.66\cdot 10^{-12}$ & $9.09\times 10^{-10}$ & $1.11 \cdot 10^{-7}$ & $1.04 \cdot 10^{-6}$ \\
        &  $2.59\cdot 10^{-12}$ & $1.97\cdot 10^{-9}$ & $1.12\cdot 10^{-7}$ & $2.52\cdot 10^{-6}$ \\
        \cmidrule(lr){2-5}

        \multirow{2}{*}{$\hat{t}_{1}$}
        & $1.11\cdot 10^4$ & $1.35\cdot 10^3$ & $5.59\cdot 10^2$ & $2.86\cdot 10^2$ \\
        & $4.33\cdot 10^3$ & $8.88 \cdot 10^2$ & $3.85 \cdot 10^2$ & $2.03 \cdot 10^2$ \\
        \cmidrule(lr){2-5}

        \multirow{2}{*}{$t_{1}$[s]}
        & $6.18\cdot 10^{13}$ & $5.70\cdot 10^7$ & $5.34\cdot 10^2$ & $2.61$ \\
        & $2.46\cdot 10^{13}$ & $1.74\cdot 10^7$ & $3.46\cdot 10^2$ & $0.77$ \\
        
        \bottomrule
    \end{tabular}
    \label{tab:Mg seg kinetics}
\end{table}


\subsubsection{ Grain boundary in Al-Mg } \label{subsubsec: STGB}

\begin{figure}[htbp]
\includegraphics[width=\linewidth]{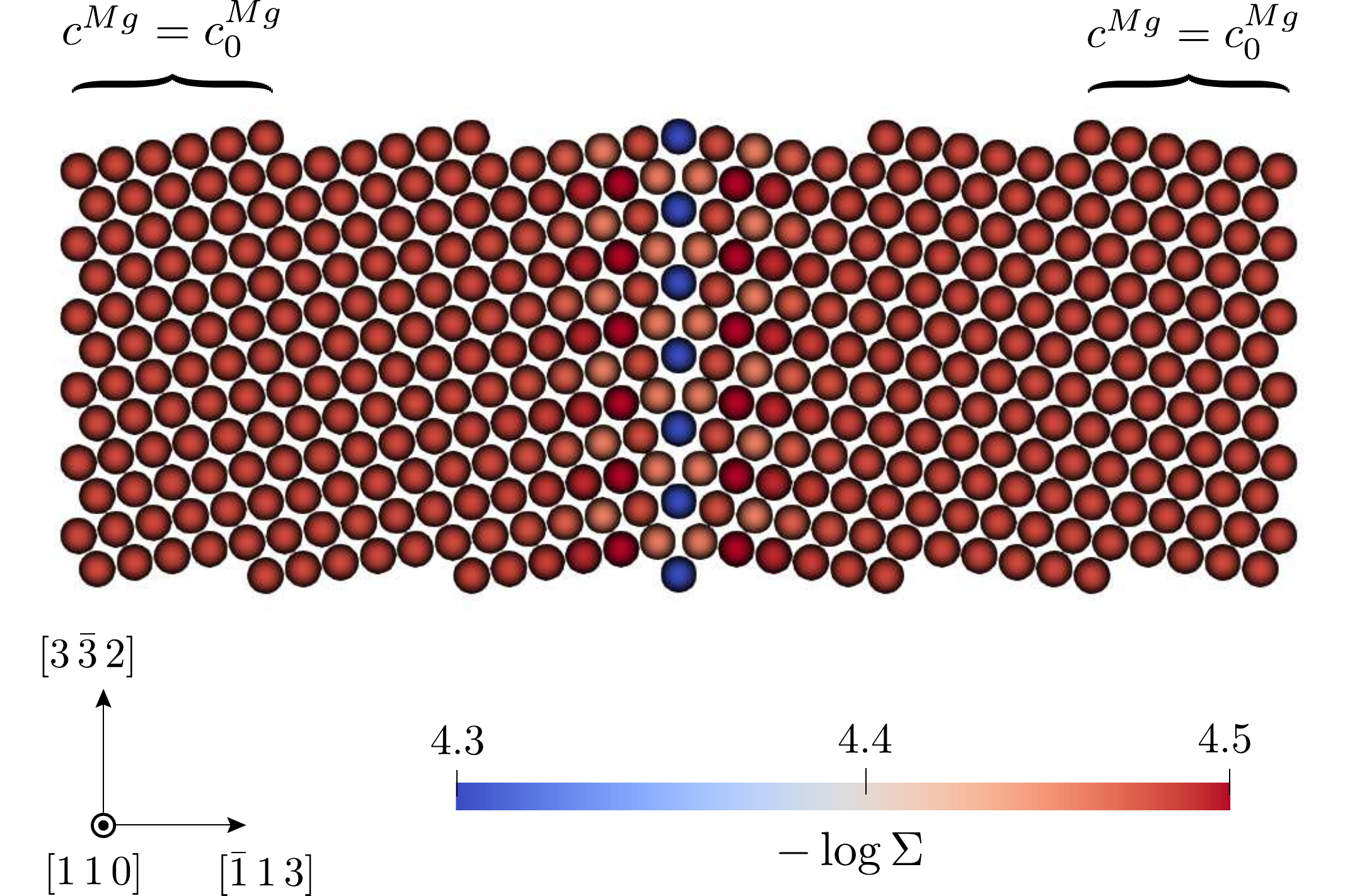}
\caption{\label{fig:GB geom} $\Sigma\,11(\bar{1}13)[110]$ grain boundary setup, color-coded by the relaxed atomic position variances at $673$~K for a uniform concentration of $c^\text{Mg} = 0.01$. Fixed concentration boundary conditions are applied at the outermost layers along the $[\bar{1}13]$ crystallographic direction, as shown. }
\end{figure}

In this section, we consider a symmetric tilt grain boundary (STGB) with rotation axis along the $[1 1 0]$ crystallographic direction and the GB plane being $(\bar{1} 1 3)$. Similar to Sec.~\ref{subsubsec: SF} (and using the same EAM interatomic potential \citep{liu1998grain}), we simulate the long-term segregation kinetics and equilibrium concentration profiles for Mg near the aforementioned GB at $673~$K for different bulk Mg concentrations. The simulation setup is shown in Fig.~\ref{fig:GB geom}, where atomic sites are color coded by the negative log of their position variances, $-\log{\Sigma_i}$, for a uniform concentration of $c^\text{Mg} = 0.01$ at all atomic sites.

\begin{figure}[htbp]
\includegraphics[width=\linewidth]{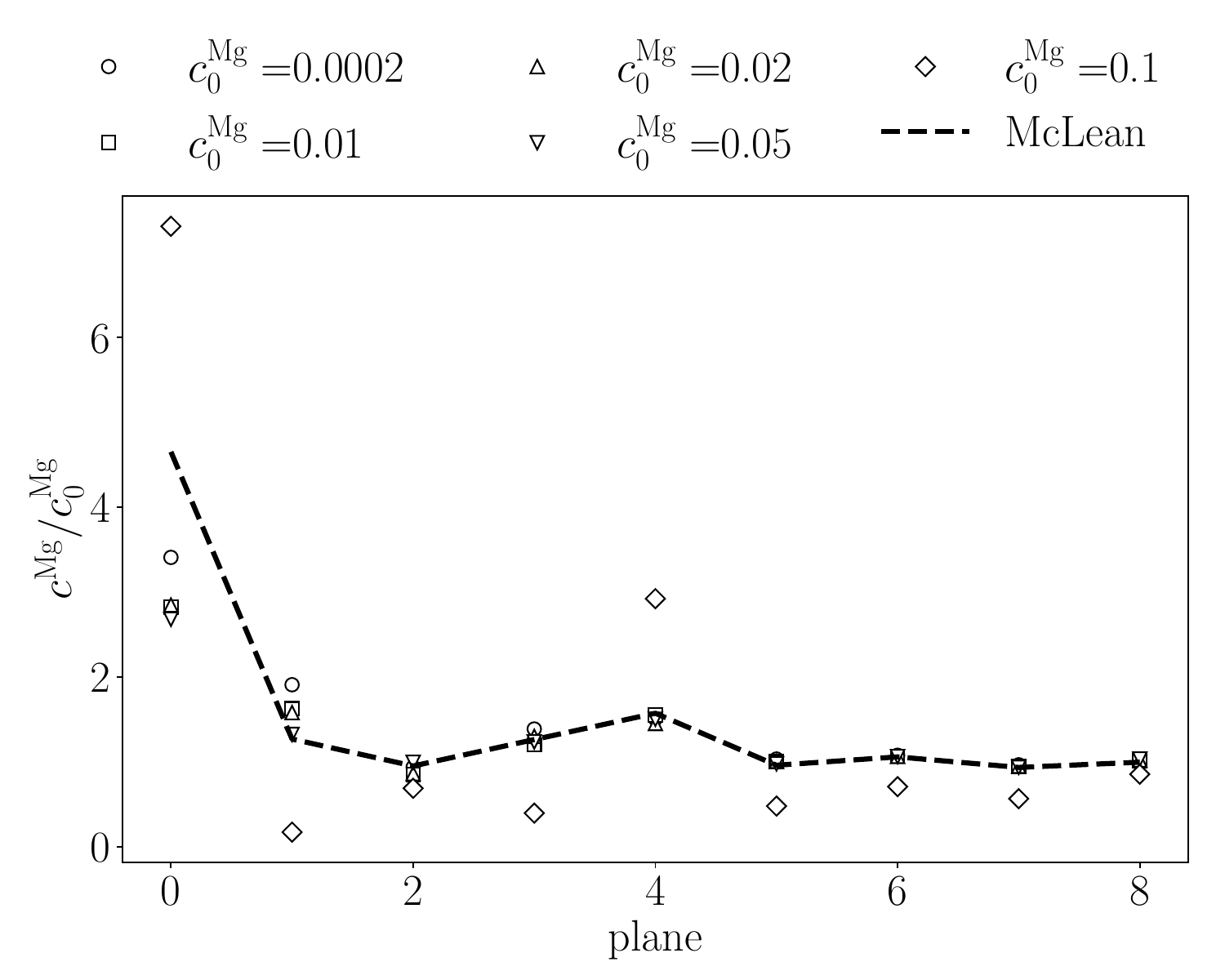}
\caption{\label{fig:GB EF equilibirum}  Enrichment factors for Mg near a $\Sigma 11\, (\bar{1} 1 3)[1 1 0]$ STGB in Al-Mg at $673~$K for different bulk Mg concentrations. }
\end{figure}

Similar to the stacking fault example, we perform long-term Mg concentration updates, using Eq.~\eqref{eq: approx master eq we use alloys} with staggered atomic relaxations according to Eq.~\eqref{eq: EOM QS}. Fig.~\ref{fig:GB EF equilibirum} shows the equilibrium Mg enrichment factors of up to 8 planes away from the GB plane for different bulk Mg concentrations. Results are compared to the Langmuir-McLean isotherm, which is valid in the dilute limit. During simulations, we do not constrain the total number of Mg atoms in the ensemble. Instead, we impose a fixed Mg concentration onto some of the outermost layers of atoms along the $[\bar{1}13]$ crystallographic direction, as shown in Fig.~\ref{fig:GB geom}. This allows Mg atoms to enter and escape the simulation geometry along this direction, thus simulating the STGB in a infinite bulk with a large reservoir of Mg atoms. The obtained enrichment factors for dilute cases tend to be closest to the McLean limit, whereas deviations are seen for increasingly non-dilute alloys. It is interesting to note that the segregation trend for $c^\text{Mg}_0 = 0.1$ (a highly non-dilute case) is drastically different than the dilute limit. This is expected due to the Al-Mg bonds being stronger than the Mg-Mg bonds \citep{DAS2023116661}. Hence, when the Mg concentration at a plane becomes too high, the next plane prefers to considerably reduce its Mg concentration. A similar oscillatory trend was observed in Fig.~6(e) of \citet{liu1998grain}, who reported the results of Monte-Carlo (MC) simulations for the same STGB using the same interatomic potential at $c_0^\text{Mg} = 0.1$. However, the total number of Mg atoms in the domain was fixed in their case, and hence their enrichment factors are lower compared to those in Fig.~\ref{fig:GB EF equilibirum} here and cannot be directly compared.

\begin{figure}[htbp]
\includegraphics[width=\linewidth]{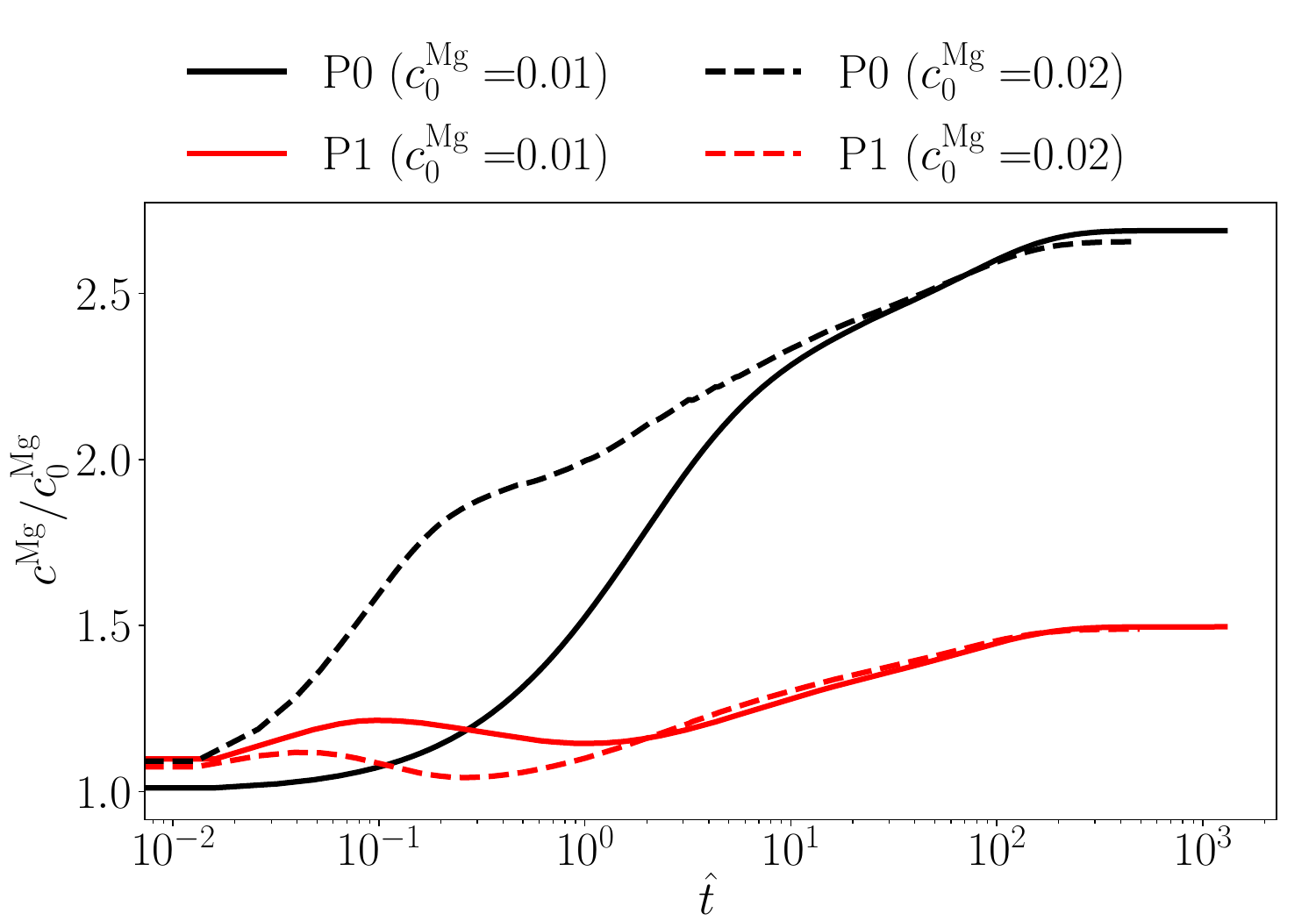}
\caption{\label{fig:GB EF kinetics}  Magnesium segregation kinetics at 673~K for different bulk magnesium concentrations at the GB plane (P0) and the adjacent plane (P1) vs.\ dimensionless time~$\hat t$. }
\end{figure}

We show the segregation kinetics for the Mg concentration evolution in the planes at and immediately next to the GB in Fig.~\ref{fig:GB EF kinetics} for $1\%$ and $2\%$ bulk Mg concentration $c_0^\text{Mg}$ vs.\ the dimensionless time $\hat{t} = c_0^v \gamma_0t$. As before, we computed the free energy of vacancy formation ($F_\text{vf}$) and the bulk vacancy concentration ($c_0^v$) at $673~$K to obtain real-time estimates for Mg to reach within $1\%$ of its equilibrium value. The reference values of the position variance $\Sigma_0$, energy barrier $E_{\text{b}_0}$, bulk vacancy concentration $c^v_0$, and dimensionless time $\hat{t}_1$ and real time $t_1$ in seconds to attain equilibrium are shown in Tab.~\ref{tab:GB Mg seg kinetics} for both values of $c_0^{Mg}$. Similar to Sect.~\ref{subsubsec: SF}, we observe that the equilibration time is shorter for a $2\%$ Mg solid solution than a $1\%$ Mg solid solution in Al.

\begin{table}[htb]
    \centering
    \caption{Magnesium segregation kinetics to the grain boundary plane for $c_0^\text{Mg} = 0.01$ and $0.02$. $t_1$ refers to the time required for the stacking fault plane to reach within $1 \%$ of its equilibrium Mg concentration.}
    \begin{tabular}{ccccc}
        \toprule
  $c_0^\text{Mg}$  & $c^v_0$ & $-\log{\Sigma_0}$ & $\hat{t}_1$ & $t_1[s]$ \\
        \midrule
        0.01 & $5.42\times10^{-6}$ & 4.48 & $2.15\times10^2$ & 0.17 \\
       0.02 & $7.08\times10^{-6}$ & 4.47 & $1.97\times10^2$ & 0.12 \\
        \bottomrule
    \end{tabular}
    \label{tab:GB Mg seg kinetics}
\end{table}

\section{Conclusion} \label{sec: conclusion}
We present a computationally efficient numerical technique to simulate the chemo-thermo-mechanically coupled substitutional diffusion of vacancies and vacancy-mediated substitutional diffusion of solutes with atomic resolution up to realistic time scales of engineering interest. The approach requires only an underlying interatomic potential and does not rely on continuum-level assumptions of diffusion kinetics. The technique relies on the separation of time scales between thermal equilibration and configurational permutations in a crystalline atomic ensemble. This permits the use of a staggered procedure with each step corresponding to: (i)~an explicit Euler time update of atomic concentrations using Harmonic Transition State Theory, followed by (ii)~a finite-temperature thermal relaxation to find equilibrium mean atomic positions and position variances for the new atomic concentrations, using the framework of Gaussian Phase Packets (GPP) \citep{gupta2021nonequilibrium}. 

We use nudged elastic band calculations of uniquely chosen atomic configurations on the fly to obtain accurate estimates of energy barriers, vacancy segregation free energies, and solute-vacancy interaction free energies, combined with the GPP-relaxed position variances to estimate the jump attempt frequency. This has been validated by comparing the bulk self-diffusion coefficient in copper to experimental data, and by comparing the self-diffusion and magnesium diffusion in aluminum near a stacking fault and a symmetric tilt grain boundary to the corresponding Langmuir-Mclean isotherms. Applying the framework to study the kinetics of vacancy and magnesium diffusion near defects in FCC aluminum shows that the simulated time scales accessible by this approach are orders of magnitude larger than those of conventional techniques such as molecular dynamics.

For vacancy-mediated solute diffusion in alloys, we present a framework normalized by a known dilute vacancy bulk concentration, which allows us to reach time scales of engineering interest. This is demonstrated by showing the magnesium segregation to a stacking fault in an Al-Mg alloy over a time span ranging from approximately 2~seconds at 600~K to $>10^{5}$ years at 300~K, assuming an equilibrium bulk vacancy concentration. This demonstrates that our atomic simulations can access intriguingly long time scales without relying on continuum modeling assumptions, while still remaining computationally tractable. The equilibrium magnesium concentration profiles have also been compared to the Langmuir-McLean isotherm, showing convincing agreement in the dilute limit.

\begin{acknowledgments}
The authors thank Prof.~William Curtin for very helpful technical discussions and Dr. Gerhard Bräunlich (ETH SIS) for supporting the numerical implementation. The support from the European Research Council (ERC) under the European Union’s Horizon 2020 research and innovation program (grant agreement no.~770754) is gratefully acknowledged.
\end{acknowledgments}

\appendix

\section{ Simplified master equation  } \label{appendix: ME simpl.}
Here we show the simplification of the general master equation for the configurational kinetics, Eq.~\eqref{eq: original master eq}, to its reduced form in Eq.~\eqref{eq: master eq 1}, which forms the basis for concentration updates in this work. This derivation has been adapted from \cite{Nastar01012000} for completeness. The concentration of a chemical species $\alpha$ at site $i$ can be written as $c^\alpha_i = \langle n_i^\alpha \rangle = \sum _{\bm n} P(\bm n,t) n^\alpha_i $. Using Eq.~\eqref{eq: original master eq}, the rate of change of this concentration can be written as
\begin{align}\label{eq:MasterEqAppendix}
    \frac{\dd c_i^\alpha }{\dd t} = \sum_{\bm n} \sum_{\tilde{\bm n}} n_i^\alpha \left[  P(\tilde{\bm n},t) W_{ \tilde{\bm n} \rightarrow \bm n} - P(\bm n,t) W_{\bm n \rightarrow \tilde{\bm n}}  \right],
\end{align}
where the sums are implied over all possible configurations $\bfn$, $\tilde\bfn$.
Noticing that the occupancy number at any site can be either 0 or 1 and that only $n_i^\alpha = 1$ contributes to the sum above, we can split the right-hand side of Eq.~\eqref{eq:MasterEqAppendix} as
\begin{align}
    \frac{\dd c_i^\alpha }{\dd t} = \sum_{\bm n \vert n_i^\alpha = 1 } \sum_{\tilde{\bm n} \vert \tilde{n}_i^\alpha = 1} n_i^\alpha \left[  P(\tilde{\bm n},t) W_{ \tilde{\bm n} \rightarrow \bm n} - P(\bm n,t) W_{\bm n \rightarrow \tilde{\bm n}}  \right] \nonumber \\ 
    + \sum_{\bm n \vert n_i^\alpha = 1 } \sum_{\tilde{\bm n} \vert \tilde{n}_i^\alpha = 0} n_i^\alpha \left[  P(\tilde{\bm n},t) W_{ \tilde{\bm n} \rightarrow \bm n} - P(\bm n,t) W_{\bm n \rightarrow \tilde{\bm n}}  \right]
    \label{eq: sum split}
\end{align}
For the first double sum in the equation above, we consider configurations $\bm n$ and $\tilde{\bm n}$ such that $n_i^\alpha = \tilde{n}_i^\alpha = 1$. Exploiting the permutation of double sums, the first term in the first double sum can be re-written as
\begin{align}
    \sum_{\bm n \vert n_i^\alpha = 1 } \sum_{\tilde{\bm n} \vert \tilde{n}_i^\alpha = 1} & \tilde{n}_i^\alpha   P(\tilde{\bm n},t) W_{ \tilde{\bm n} \rightarrow \bm n} = \nonumber \\
    &\sum_{\tilde{\bm n} \vert \tilde{n}_i^\alpha = 1} \sum_{\bm n \vert n_i^\alpha = 1 }  n_i^\alpha   P(\bm n,t) W_{ \bm n \rightarrow \tilde{\bm n}},
\end{align}
which is equal to the second term in the first double sum in Eq.~\eqref{eq: sum split}. Hence, only the second double sum contributes to the rate of concentration change in Eq.~\eqref{eq: sum split}. This is intuitive, because a species change at site $i$ happens only if the new configuration $\tilde{\bm n}$ is such that $\tilde{n}_i^\alpha = 0$. In other words, site $i$ is occupied by some other chemical species or a vacancy in configuration $\tilde{\bm n}$. If the new occupancy at site $i$ corresponds to another chemical species, this can be achieved only by a coordinated movement of atoms. Such a coordinated movement might be due to a coordinated exchange of atoms without involving any vacancy (e.g., ring-based mechanisms \citep{mehrer2007diffusion}). However, the Kirkendall experiments \citep{kirkendall1939rates} have shown that vacancies play an important role in substitutional diffusion in crystalline metals. Other types of coordinated atomic movements might be due to long chains of moving atoms in a disordered region such as grain boundaries \citep{ChesserKojuMishin+2024+85+105}. However, we restrict ourselves to a nearest-neighbor vacancy-mediated diffusion mechanism for simplicity. This implies that the configuration $\bm n$ corresponds to a vacancy at any of the nearest-neighbors to site $i$. Therefore, assuming that configurations $\bm n$ and $\tilde{\bm n}$ differ only by the exchange of atom $\alpha$ from site $i$ to any nearest-neighbor vacant site $j$, the second sum in Eq.~\eqref{eq: sum split} can be re-written as
\begin{align}
    \frac{\dd c_i^\alpha }{\dd t} &= \sum_{\bm n} P(\bm n,t) \sum_{j \in \text{NN}^i} \left[ n_i^v n_j^\alpha \gamma_{j \rightarrow i}^{\alpha} - n_j^v n_i^\alpha \gamma_{i \rightarrow j}^{\alpha}     \right] \nonumber \\
   &= \sum_{j \in \text{NN}^i} \left \langle n_i^v n_j^\alpha \gamma_{j \rightarrow i}^{\alpha} - n_j^v n_i^\alpha \gamma_{i \rightarrow j}^{\alpha}    \right \rangle ,
    \label{eq: app last}
\end{align}
where $\gamma_{j \rightarrow i}^{\alpha} = W_{ \tilde{\bm n} \rightarrow \bm n} $ and $\gamma_{i \rightarrow j}^{\alpha} = W_{ \bm n \rightarrow \tilde{\bm n}} $ are the transition rates of the system corresponding to this atomic hop to a nearest-neighbor vacancy. With this, we have arrived at Eq.~\eqref{eq: master eq 1}, which is used in this work.

\bibliography{substitutional_mass_diffusion}

\end{document}